\documentclass[pre,aps,twocolumn,superscriptaddress,showpacs,floatfix]{revtex4}

\usepackage{epsfig,graphicx,latexsym}
\usepackage{amssymb,amsmath}

\newcommand{\ve}[1][K]{\mathbf{#1}}

\begin{document}
\author{T. Gu\'erin}
\author{D. S. Dean}
\affiliation{Laboratoire Ondes et
Mati\`ere d'Aquitaine (LOMA), CNRS, UMR 5798 / Universit\'e de  Bordeaux, F-33400 Talence, France}

\begin{abstract}
We consider the  dispersion properties of tracer particles moving in non-equilibrium heterogeneous periodic media. The tracer motion is described by a Fokker-Planck equation with arbitrary spatially periodic (but constant in time) local diffusion tensors and drift, eventually with the presence of obstacles. 
We derive a Kubo-like formula for the time dependent effective diffusion tensor valid in any dimension.
From this general formula, we derive expressions for  the late time effective diffusion tensor and drift in these systems. In addition, we find an explicit formula for the late finite time corrections to these transport coefficients. In one dimension, we give closed analytical formula for the transport coefficients. The formulas derived here are very general and provide a straightforward method to compute the dispersion properties in arbitrary non-equilibrium periodic advection-diffusion systems.
\end{abstract}

\title{Kubo formulas for dispersion in heterogeneous periodic non-equilibrium  systems}
\maketitle

\section{Introduction}
Fokker-Planck (FP) equations, and their associated stochastic differential equations (SDE) arise in a huge variety of physical systems~\cite{VanKampen1992,oksendal2003stochastic,gardiner1983handbook}. In fluid mechanics and hydrology, such equations describe the transport of passive scalars advected by a fluid and locally dispersed by a microscopic molecular diffusivity. They also describe the diffusion of colloids and polymers in soft matter physics, as well as charge and energy transport in solid state physics. In many systems, damping due to the environment is very strong and inertial effects are thus negligible. In such systems, the evolution of the probability density function (pdf) $p(\ve[x],t)$ of a tracer particle at position $\ve[x]$ and time $t$ obeys the FP equation 
\begin{equation}
\partial_t p=\partial_{x_i}\left\{\partial_{x_j}[\kappa_{ij}({\bf x})p]- u_i({\bf x}) p\right\}=-H_{\ve[x]} p,\label{fp}
\end{equation}
where the Einstein convention on indexes summation is used, and the indexes $i,j$ run from $1$ to the spatial dimension $d$. 
The time independent tensor $\kappa_{ij}({\bf x})$ is the local diffusivity tensor and can, in general, vary in space in media of non-uniform composition, since diffusivity is depends on the local material properties of the system. The local diffusivity may also vary in space due to temperature inhomogeneities, or in the case of particle diffusion in viscous fluid flows close to solid boundaries \cite{brenner1961slow}. The term $u_i({\bf x})$ represents a drift term generated by forces such as buoyancy forces or electromagnetic forces acting on colloids, fluid flows, as well as thermodynamic effects, such as thermodiffusion in the presence of temperature  gradients \cite{wurger2010thermal}.  The presence of impenetrable obstacles can also be treated by introducing surfaces at which there is no normal flux, 
\begin{align}
	n_i({\bf x})\{\partial_{x_j}[\kappa_{ij}({\bf x})p]- u_i({\bf x}) p\}=0, \hspace{0.5cm} \ve[x]\in S,
\end{align}
where $n_i$ represents the normal vector to the surface $S$ of the obstacles. 

Equation (\ref{fp}) describes the  motion of tracer particles at small scales in a wide range of heterogeneous media. At a coarse-grained level, the motion is described by effective transport coefficients characterizing the average flux of particles and the relative spreading with time of initially close particles. From a theoretical point of view, the dispersion of tracer particles is defined  in terms of two quantities, the mean  displacement $\mu_i(t)$ during a time $t$ in the direction $i$ 
\begin{equation}
 \mu_i(t) = \langle X_i(t) -X_i(0)\rangle, \label{DefMu_i}
\end{equation}
(where ${\bf X}(t)$ denotes the position of a tracer particle and $\langle \cdot\cdot\cdot\rangle$ denotes ensemble averaging), 
and the correlation of the dispersion in the directions $i$ and $j$ defined by
\begin{equation}
\sigma_{ij}(t) = \langle [X_i(t) -X_i(0)][X_j(t) -X_j(0)]\rangle-\mu_i(t)\mu_j(t). \label{DefSigma}
\end{equation}
The quantity $\sigma_{ij}(t)$ thus characterizes the dispersion of a cloud of particles  about its mean position. Of particular interest  is the large time behavior of the dispersion properties, 
\begin{align}
&\mu_i(t)\simeq V_i \ t, \hspace{1.5cm}&(t\rightarrow\infty)&\\
&\sigma_{ij}(t)\simeq 2 \ t\ (D_{ij}  + C_{ij}/t), \hspace{0.5cm}&(t\rightarrow\infty)&, \label{DefLateTimeCoeff}
\end{align}
where $V_i$ is the average particle drift, $D_{ij}$ is the late time effective diffusion tensor, and $C_{ij}$ gives the leading order corrections and provides information on how fast the system approaches the diffusive limit. We will see that the asymptotic form (\ref{DefLateTimeCoeff}) is valid if the medium 
is spatially periodic. The quantities $V_i$ and $D_{ij}$  characterize the late time, large scale transport properties of the system and are routinely measured in experimental systems  by 
single particle tracking methods \cite{daumas2003confined} or by ensemble based measurements such as fluorescence recovery after photobleaching (FRAP) \cite{reits2001fixed}. Effective transport coefficient are also important  for estimating the spread of pollutants, filtration, mixing and chemical reaction times \cite{condamin2007}. 

Transport and dispersion properties in non-equilibrium periodic systems have been investigated at length in various contexts. In a pioneering, classical work, Taylor~\cite{taylor1953dispersion} considered the dispersion of particles moving in viscous flows in cylindrical containers, and showed that the effective diffusion coefficient  could be orders of magnitudes larger than the molecular diffusivity. Effective transport in incrompressible fluid flows were also investigated at length, such as in the case of diffusion in Rayleigh-B\'enard convection cells \cite{rosenbluth1987effective,Shraiman1987,McCarty1988}, in frozen turbulent flows \cite{majda1999simplified} or in porous media \cite{brenner1980dispersion,rubinstein1986dispersion,quintard1994convection,alshare2010modeling,souto1997dispersion,souto1997dispersion1,quintard1993transport}
using homogenization theory. However, these approaches have been restricted so far to incompressible flows and it is not clear how to generalize them. In another context, results were also derived, with the methods of statistical physics, for particles diffusing in periodic potentials when the diffusivity is constant \cite{dean2007effective,derrida1983velocity,zwanzig1988diffusion,deGennes1975brownian,lifson1962self}; in here the potential barrier slows down the diffusion as the particle tends to become trapped in local mining. The non-equilibrium problem of transport in one-dimensional  potential in the presence of external constant force (still with uniform microscopic diffusivity) was also considered \cite{reimann2001giant,lindner2001optimal,reimann2002diffusion,reimann2008weak}; this system is predicted to exhibit a huge increase of the effective diffusion near a critical tilting force. Results have also been obtained in the more general case where the noise amplitude is a periodic function of  position \cite{lindner2002,reguera2006entropic,burada2008entropic}, notably in the context of the modeling of transport in periodic channels within the Fick-Jacobs approximation (where the full Fokker-Planck equation is approximated by an effective one dimensional one). In many of the  above cases, results  only exist in one dimension, however their generalization to higher dimensions is implicit in the general results we will present here.

In this paper, we present a general formalism to compute the time-dependent dispersion tensor $\sigma_{ij}(t)$ (section \ref{gk}), the late  time diffusion tensor $D_{ij}$ (section \ref{eft}) and the corrections $C_{ij}$ describing the approach to the diffusive limit (section  \ref{late}). A summary of the results is given in section \ref{SummaryResults} for the reader who might not be interested in the technical aspects of the calculations. The results derived are valid for any system described by Eq. (\ref{fp}) for periodic $\kappa_{ij}(\ve[x])$ and $u_i(\ve[x])$; they encapsulate all known results for this problem and also generalize them to more complex situations. We will explicitly compare them to the expressions of the literature in various cases (section \ref{SectionSpecialCases}), and we will also give an explicit fully analytical solution for the one dimensional problem (section \ref{1d}). 
A brief derivation of our main results has already appeared in Ref.~\cite{guerin2015}, where they were used to analyze diffusion in a system with a spatially varying diffusivity in the presence of a uniform applied force. The present paper extends this recently developed approach, new features are the computation of the time-dependent dispersion properties (instead of only the late time diffusive limit) and that the incorporation of the effect of impenetrable obstacles with no-flux boundary conditions. Furthermore, the derivation presented here is different from that appearing in Ref.~\cite{guerin2015}, as it is not based on the SDE formulation used there. Finally, the present paper contains  explicit expressions for the first temporal corrections $C_{ij}$ to the late time diffusion tensor. We will show that the $1/t$ decay of the late time  correction  [Eq. (\ref{DefLateTimeCoeff})] occurs in all dimensions. This universality can be understood as, for periodic systems, the Fokker-Planck equation restricted to the basic cell $\Omega$ will in general have a gap and not a continuum (whose density would be dependent on the spatial dimension) of eigenvalues near zero.

\section{Notations and summary of results}
\label{SummaryResults}
We assume that the space can be divided into individual cells $\Omega$, and we  that the diffusivity tensor $\kappa_{ij}(\ve[x])$ and the drift $u_i(\ve[x])$ are periodic in space, and thus satisfy
\begin{align}
&	\kappa_{ij}(\ve[x])=\kappa_{ij}(\ve[x]+\ve[k]),\hspace{0.8cm}	u_{i}(\ve[x])=u_i(\ve[x]+\ve[k]), \label{PeriodProperty}
\end{align} 
for all vectors $\ve[k]$ joining the centers of the cells composing the periodic structure. Denoting by $\ve[X](t)$ the position of an individual particle at time $t$, we define the propagator $p(\ve[x],t\vert\ve[y])$ of the stochastic process $\ve[X](t)$ in infinite space; $p(\ve[x],t\vert\ve[y])$ is thus  the probability density to observe a particle at position $\ve[x]$ at time $t$ starting from $\ve[y]$ at time $0$. Mathematically, $p$ is the solution of the equation in infinite space
\begin{align}
&(\partial_t+H_{\ve[x]})p(\ve[x],t\vert \ve[y])=0 \ ; \hspace{0.4cm}
p(\ve[x],0\vert\ve[y])=\delta(\ve[x]-\ve[y])\label{EqPropInfSpace}.
\end{align}
We also introduce the process $\tilde{\ve[X]}(t)$ as the  position of the particle $\ve[X]$ "modulo $\Omega$", $\tilde{\ve[X]}$ is defined as the position of the particle in a reference cell $\Omega_0$ obtained after an integer number of translations of $\ve[X]$ along the vectors joining the centers of the cells composing the periodic structure. We introduce the propagator $P(\ve[x],t\vert\ve[y])$ for the process modulo $\Omega$, $P$ thus represents the probability density, starting at position $\ve[y]$ at time $0$, to be at time $t$ at position $\ve[x]$ modulo $\Omega$. $P$ is the solution of the FP equation (\ref{fp}) with periodic boundary conditions on the boundaries of $\Omega$ and initial value $P(\ve[x],0\vert\ve[y])=\delta(\ve[x]-\ve[y])$.  There is an obvious relation between these propagators,
\begin{align}
P(\ve[x],t\vert\ve[y])=\sum_{\ve[k]} \ p(\ve[x]+\ve[k],t \vert\ve[y]) ,\label{LinkPropag}
\end{align}
where $\ve[k]$ represents the vectors joining the center of the cell containing $\ve[x]$ to the centers of all the other cells of the periodic structure. The value of $P$ at late times is denoted by  $P_0$ and is the steady-state distribution of particles in one period (modulo translations along the lattice vectors). In contrast, the propagator in infinite space $p$ does not tend to a steady state, but becomes Gaussian at long times, with a variance growing linearly as $t$, representing the spreading of the particle density. In what follows, we always assume that at $t=0$ we are in the steady state, that is to say that the process $\tilde{\ve[X]}(t=0)$ modulo $\Omega$ has the distribution $P_0$.  

Before entering the details of the calculations, let us give briefly our main results. First, we will show that the temporal Laplace transform $\hat\sigma_{ij}(s)=\int_0^{\infty}\sigma_{ij}(t)e^{-st}dt$ of the dispersion tensor is given by
\begin{align}
& \hat\sigma_{ij}(s)=\frac{2}{s^2} \int_\Omega d{\bf x}\  \kappa_{ij}({\bf x})P_0({\bf x})\nonumber\\
 & -\frac{1}{s^2}\iint_{\Omega}d\ve[x]d\ve[y][\tilde u_i(\ve[x])\tilde u_j^*(\ve[y])+\tilde u_j(\ve[x])\tilde u_i^*(\ve[y])]\hat P'(\ve[x],s\vert\ve[y])P_0(\ve[y]) \label{9553}.
\end{align}
In this relatively compact formula, $\hat P'(\ve[x],s\vert\ve[y])$ represents the Laplace transform of $P(\ve[x],t\vert\ve[y])-P_0(\ve[x])$. We have also defined an effective velocity field $\tilde u$ that takes into account the presence of obstacles (if present)
\begin{align}
	\tilde u_i(\ve[x])=u_i(\ve[x])- n_j(\ve[x]) \kappa_{ij}(\ve[x])\delta_S(\ve[x]),
\end{align}
where $\delta_S$ is a surface delta function; to be clear with the notations we precise that 
\begin{align}
	\int_{\Omega} d\ve[x]  n_j(\ve[x]) \kappa_{ij}(\ve[x])\delta_S(\ve[x])\psi(\ve[x])=\int_{S_{\Omega}} dS_j \kappa_{ij}(\ve[x]) \psi(\ve[x])
\end{align}
for any function $\psi$, with the infinitesimal normal surface vector $dS_i$ and the unit normal vector $n_i$ oriented towards the interior of the obstacles. Finally, we have also introduced the drift field $u_i^*$ which can be interpreted as the drift field for the particles upon time reversal, and is given by
\begin{align}
u_i^*(\ve[x])=u_i(\ve[x])-2\frac{J_{i}(\ve[x])}{P_0(\ve[x])}, 
\end{align}
where we have introduced the current in the steady state $J_{i}$ 
\begin{align}
	J_{i}(\ve[x])=u_i(\ve[x])P_0(\ve[x])-\partial_{x_k}[\kappa_{ik}(\ve[x])P_0(\ve[x])].
\end{align} 
The notation $\tilde u_i^*(\ve[x])$ denotes that we have added surface drift terms which prevent the particles from entering the obstacles,
\begin{align}
	\tilde u_i^*(\ve[x])=u_i^*(\ve[x])- n_j(\ve[x]) \kappa_{ij}(\ve[x])\delta_S(\ve[x]).
\end{align}
Equation (\ref{9553}) states clearly that the large scale dispersion properties can be obtained from the properties of the propagator (with periodic boundary conditions), which is used to average the drift fields with and without time-reversal; this is the first result of the paper. 

The second main result is an expression for the late time diffusion tensor $D_{ij}$. Taking the large time limit of  Eq.~(\ref{9553})  leads to
\begin{align}
D_{ij}=&\int_\Omega d{\bf x}\  \kappa_{ij}({\bf x})P_0({\bf x})\nonumber\\
&+\frac{1}{2}\int_{\Omega}d\ve[x]\ [ u_i(\ve[x])f_j(\ve[x])+ u_j(\ve[x])f_i(\ve[x])]\nonumber\\
&-\frac{1}{2}\int_{S_\Omega}dS_l(\ve[x])\ [\kappa_{il}(\ve[x])f_j(\ve[x])+ \kappa_{jl}(\ve[x])f_i(\ve[x])],\label{KuboDij}
\end{align}
where the function $f_j$ is defined by
\begin{align}
	f_j(\ve[x])=
 &-\int_{\Omega}d\ve[y]\  P_0(\ve[y])G(\ve[x] \vert\ve[y]) \tilde u_j^*(\ve[y]) 
\end{align}
and 
\begin{align}
G(\ve[x]\vert\ve[y])=\int_0^{\infty}dt[P(\ve[x],t\vert\ve[y])-P_0(\ve[x])].
\end{align}
$G$ is the pseudo-Green function \cite{barton1989elements} of the operator $H$. For computational purposes, it is useful to note that $f_i$ can be obtained by solving the non-homogeneous linear partial differential equation 
\begin{align}
H_{\ve[x]}f_i(\ve[x]) =&  - u_i^*({\bf x})P_0({\bf x}) + P_0(\ve[x]) \int_\Omega d{\bf y} \ 
 u_i^*(\ve[y])P_0({\bf y})
 \nonumber\\
 &-P_0(\ve[x])\int_{S_{\Omega}}dS_l(\ve[y]) P_0(\ve[y]) \kappa_{li}(\ve[y]) \label{eqfiIntro}
\end{align}
with periodic boundary conditions and also with the condition
\begin{align}
	n_j\{u_j({\bf x}) f_i(\ve[x])-\partial_{x_k}[\kappa_{kj}({\bf x})f_i(\ve[x])]\}&=-n_j \kappa_{ji}(\ve[x])P_0(\ve[x]) 
\end{align}
at the surface of the obstacles, together with the orthogonality condition
\begin{equation}
	\int_\Omega d{\bf x} \ f_i({\bf x})=0.  
\end{equation}
The above expressions are exact, general and can be used to compute the dispersion properties in any advection-diffusion problem with periodic diffusivity tensor and drift fields, in any dimension, in the presence of obstacles with reflecting boundaries.   We will show that our expression for $D_{ij}$ contains existing expressions for the effective diffusion tensor in porous media for particles transported by incrompressible fluids \cite{brenner1980dispersion,carbonell1983dispersion}, diffusion in periodic potentials \cite{drummond1987effective}, and diffusion in non-equilibrium one dimensional systems  \cite{reimann2001giant,reimann2002diffusion,burada2008entropic}.

\begin{figure}%
\includegraphics[width=7cm]{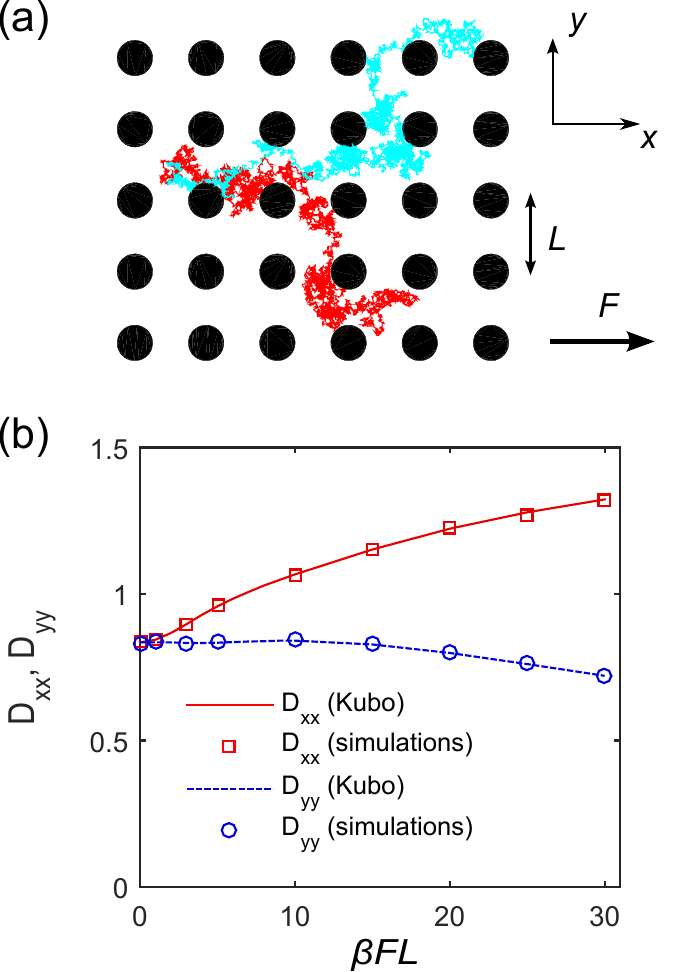}%
\caption{(a) Examples of stochastic trajectories of two tracer particles diffusing in a heterogeneous medium (here a 2D array of circular obstacles) under a force $F$ (arrow) that biases the motion to the right. (b) Elements of the dispersion tensor for the system described in (a), calculated from our theoretical formulas Eqs.~(\ref{KuboDij},\ref{eqfiIntro}) (full and dashed lines)  \textit{vs} results of stochastic simulations (symbols). Parameters: disk radius $R=0.25L$ with $L$ the length of the side of the periodic cell. Results for $D_{ij}$ are expressed in units of the (constant) diffusion coefficient of the tracers outside the obstacles.}
\label{figForceObstacles}%
\end{figure}

One may also derive find an alternative expression for $D_{ij}$ by defining a function $f_i^*$ as 
\begin{align}
	f_i^*(\ve[y])=
 &-\int_{\Omega}d\ve[x]\  P_0(\ve[y])G(\ve[x] \vert\ve[y]) \tilde u_i(\ve[x]),  
\end{align}
in terms of which the effective diffusion tensor reads 
\begin{align}
D_{ij}=&\int_\Omega d{\bf y}\  \kappa_{ij}({\bf y})P_0({\bf y})\nonumber\\
&+\frac{1}{2}\int_{\Omega}d\ve[y]\ [u_i^*(\ve[y])f_j^*(\ve[y])+u_j^*(\ve[y])f_i^*(\ve[y])] \nonumber\\
&- \frac{1}{2}\int_{S_{\Omega}}dS_l(\ve[y])[\kappa_{il}({\bf y})f_j^*(\ve[x])+\kappa_{jl}({\bf y})f_i^*(\ve[y])].
\end{align}
This function $f_i^*$ can also be obtained as the solution of the partial differential equations
\begin{align}
H_{\ve[y]}^*f_i^*(\ve[y])=&  - u_i({\bf y})P_0({\bf y}) + P_0(\ve[y]) \int_\Omega d{\bf x} \ 
[u_i(\ve[x])P_0({\bf x})]
 \nonumber\\
 &-P_0(\ve[y])\int_{S_{\Omega}}dS_l(\ve[x]) P_0(\ve[x]) \kappa_{li}(\ve[x]).\label{eqgi}
\end{align}
where we have denoted $H^*$ the transport operator for the time-reversed process, 
\begin{align}
H_{\ve[y]}^*(\cdot)=\partial_{y_i}[u_i^*(\ve[y])(\cdot)]-\partial_{y_i}\partial_{y_k}[\kappa_{ij}(\ve[y])(\cdot)]. 
\end{align}
The function $f_i^*$ has periodic boundary conditions, and satisfies at the surface of the obstacles
\begin{align}
	n_j({\bf y})\{u_j^*({\bf y}) f_i^*(\ve[y])-\partial_{y_k}[\kappa_{kj}({\bf y})f_i^*(\ve[y])]\}&=-n_j \kappa_{ji}(\ve[y])P_0(\ve[y]),
\end{align}
and also satisfies the orthogonality condition
\begin{equation}
	\int_\Omega d{\bf y} \ f_i^*({\bf y})=0.  
\end{equation}
Thus, we have two formulas for the effective diffusion tensor, which can be straightforwardly evaluated numerically by solving a set of elliptic partial differential equation on the domain $\Omega$. 

Finally, we will show that the tensor $C_{ij}$ describing the approach to the diffusive limit can be obtained from
\begin{align}
C_{ij}=&\frac{1}{2} \int_{\Omega}d\ve[x]\ \frac{f_i(\ve[x])f_j^*(\ve[x])+f_j(\ve[x])f_i^*(\ve[x])}{P_0(\ve[x])}.
\end{align}
Although numerous expressions exist in the literature for $D_{ij}$ for particular cases, the correction tensor $C_{ij}$ has to our knowledge only been studied for diffusion in a periodic potential \cite{dean2014approach} or with periodic diffusivity \cite{Dean2014PRE}, the above expressions enable its computation in a systematic way and at a computational cost equivalent to that required to calculate $D_{ij}$.    

Our results can be applied to study tracer particles diffusing in a 2D array of circular impenetrable obstacles (hard disks)  [Fig.\ref{figForceObstacles}(a)] and subjected to an external force. For this system, the elements of the effective diffusion tensor match perfectly with the results of stochastic simulations (the algorithm is presented in appendix \ref{AppSimu}).

\section{Derivation of the Kubo formula for the time dependent dispersion}
\label{gk}
Here we describe the derivation of the Kubo formula for the dispersion tensor.  We first briefly compute the average displacement in the direction $i$ during a time $t$, defined in Eq. (\ref{DefMu_i}), which is given by
\begin{align}
	\mu_i(t)=\int_{\mathbb{R}^d}d\ve[x]\int_{\Omega}d\ve[y]\ P_0(\ve[y])p(\ve[x],t\vert\ve[y])(x_i-y_i). 
\end{align}
Note that the integral over the whole space $\mathbb{R}^d$ denotes the integral over the volume not occupied by obstacles (where $p$ does not vanish).  Taking the time derivative of the above expression, and using Eq.~(\ref{fp}), we obtain
\begin{align}
\partial_t \mu_i(t) =\int_{\mathbb{R}^d}&d\ve[x]\int_{\Omega}d\ve[y]P_0(\ve[y])\{-\partial_{x_j}[u_j(\ve[x])p(\ve[x],t\vert \ve[y])]\nonumber\\
&+\partial_{x_j}\partial_{x_k}[\kappa_{kj}(\ve[x])p(\ve[x],t\vert \ve[y])]\}(x_i-y_i). 
\end{align}
Integrating by parts over $\ve[x]$ and applying the divergence theorem  yields
\begin{align}
\partial_t\mu_i(t)=\int_{\mathbb{R}^d}d\ve[x]\int_{\Omega}d\ve[y]P_0(\ve[y])\{u_i(\ve[x])p(\ve[x],t\vert\ve[y])&\nonumber\\
-\partial_{x_k}[\kappa_{ik}(\ve[x])p(\ve[x],t\vert\ve[y])]\},& 
\end{align}
where we have taken into account the reflecting boundary conditions at the obstacles, so that no surface integral has appeared.
Decomposing the integral over $\ve[x]$ in integrals over each unit cell, using the periodicity property (\ref{PeriodProperty}) of $u_i$ and $\kappa_{ij}$ and the definition (\ref{LinkPropag}), we get
\begin{align}
\partial_t\mu_i(t)=\int_{\Omega}d\ve[x]\int_{\Omega}d\ve[y]P_0(\ve[y])\{u_i(\ve[x])P(\ve[x],t\vert\ve[y])&\nonumber\\
-\partial_{x_k}[\kappa_{ik}(\ve[x])P(\ve[x],t\vert\ve[y])]\}.& \label{90584}
\end{align}
We recall that, since $P_0$ is the steady state particle distribution, it satisfies
\begin{align}
\int_{\Omega}d\ve[y]P_0(\ve[y])P(\ve[x],t\vert\ve[y])=P_0(\ve[x])\label{StStProp}.
\end{align} 
Using this property, and again applying the divergence theorem, (\ref{90584}) becomes
\begin{align}
\partial_t\mu_i(t)=&\int_{\Omega}d\ve[x] \ u_i(\ve[x])P_0(\ve[x])\nonumber\\
&-\int_{S_{\Omega}}dS_j(\ve[x]) \kappa_{ij}(\ve[x])P_0(\ve[x])\}=V_i, \label{ResultMu_i0}
\end{align}
where $S_{\Omega}$ represents the surface of the obstacles inside the cell $\Omega$, and the infinitesimal surface vector $dS_j$ is oriented towards the interior of the obstacles.  Eq.~(\ref{ResultMu_i0}) shows that, when the initial conditions are those of the steady state, the average drift $V_i$ is constant in time. This is in agreement with the classic result of Stratonovich~\cite{stratonovich1958oscillator}.

Let us now derive a Kubo formula for the time dependent dispersion tensor $\sigma_{ij}$ [defined in Eq.~(\ref{DefSigma})]. Define $\psi_{ij}$ to be
\begin{align}
	&\sigma_{ij}(t)=\psi_{ij}(t)-\mu_i(t)\mu_j(t) \label{LinkSigmaPsi}.
\end{align}
By definition, the expression for $\psi_{ij}$ is
\begin{align}
	&\psi_{ij}(t)=\int_{\mathbb{R}^d}d\ve[x]\int_{\Omega}d\ve[y]P_0(\ve[y])p(\ve[x],t\vert\ve[y])(x_i-y_i)(x_j-y_j). \label{594}
\end{align}
We denote by $\hat \phi(s)=\int_0^{\infty}dt \phi(t) e^{-st}$ the temporal Laplace transform of any function $\phi$. The Laplace transform of the FP equation (\ref{fp}) is given by
\begin{align}
s &\hat p(\ve[x] ,s\vert\ve[y])  =\delta(\ve[x]-\ve[y]) \nonumber\\
&+\partial_{x_i}\left\{\partial_{x_j} [\kappa_{ij}({\bf x})\hat p(\ve[x],s\vert\ve[y]) ]- u_i({\bf x})\hat p(\ve[x],s\vert\ve[y]) \right\}.
\end{align}
Using the above equality, the expression for $\psi_{ij}$ becomes
\begin{align}
 	\hat{\psi}_{ij}(s)&=\frac{1}{s}\int_{\mathbb{R}^d}d\ve[x]\int_{\Omega}d\ve[y] \ P_0(\ve[y])(x_i-y_i)(x_j-y_j)\nonumber\\ &\partial_{x_k}\left\{\partial_{x_l} [\kappa_{kl}({\bf x})\hat p(\ve[x],s\vert\ve[y]) ]- u_k({\bf x}) \hat p(\ve[x],s\vert\ve[y]) \right\}. \label{094}
\end{align}
We remark that
\begin{align}
	\partial_{x_k}[(x_i-y_i)(x_j-y_j)]= & \ \delta_{ki}(x_j-y_j)\nonumber\\
&+\delta_{kj}(x_i-y_i),
\end{align}
so that the integration by parts over $\ve[x]$ in Eq.~(\ref{094}) leads to an expression of the form
\begin{align}
	\hat{\psi}_{ij}(s)=\hat{E}_{ij}(s)+\hat{E}_{ji}(s)
\end{align} 
with
\begin{align}
\hat{E}_{ij}(s)=&\frac{1}{s}\int_{\mathbb{R}^d}d\ve[x]\int_{\Omega}d\ve[y] \ P_0(\ve[y])(x_j-y_j)\nonumber\\  &\left\{u_i({\bf x}) \hat p(\ve[x],s\vert\ve[y])-\partial_{x_k} [\kappa_{ki}({\bf x})\hat p(\ve[x],s\vert\ve[y]) ]  \right\}.
\end{align}
Integrating by parts the term containing $\kappa_{ik}$ leads to
\begin{align}
\hat{E}_{ij}(s)&=\frac{1}{s}\int_{\mathbb{R}^d}d\ve[x]\int_{\Omega}d\ve[y] \ P_0(\ve[y])(x_j-y_j)u_i({\bf x})\hat p(\ve[x],s\vert\ve[y]) \nonumber\\
&-\frac{1}{s}\int_{S}dS_l(\ve[x])\int_{\Omega}d\ve[y] P_0(\ve[y]) \kappa_{il}({\bf x})(x_j-y_j)\hat p(\ve[x],s\vert\ve[y])\nonumber\\
&+\frac{1}{s}\int_{\mathbb{R}^d}d\ve[x]\int_{\Omega}d\ve[y] P_0(\ve[y]) \hat p(\ve[x],s\vert\ve[y])\kappa_{ij}({\bf x}).\label{8583}
\end{align}
Using the periodicity property of $\kappa$ and the Laplace transform of the steady state property (\ref{StStProp}), we obtain
\begin{align}
	\int_{\mathbb{R}^d}d\ve[x]\int_{\Omega}d\ve[y] &P_0(\ve[y]) \hat p(\ve[x],s\vert\ve[y])\kappa_{il}({\bf x})=\nonumber\\
& s^{-1}\int_{\Omega}d\ve[x]\  P_0(\ve[x]) \kappa_{il}({\bf x})=s^{-1}\langle\kappa_{ij}\rangle_0,
\end{align}
where $\langle\cdot\cdot\cdot\rangle_0$ represents the average over $P_0$. 
Decomposing the integral appearing in (\ref{8583}) over $\ve[x]$ on all the individual cells of the periodic structure, and changing of variable  $\ve[x]\rightarrow\ve[x]+\ve[k]$, $\ve[y]
\rightarrow\ve[y]+\ve[k]$ in each of them (where $\ve[k]$ is the lattice vector such that $\ve[x]+\ve[k]$ is inside the cell $\Omega$), and summing over all lattice vectors $\ve[k]$ again, we see that we can exchange the integration domains between $\ve[x]$ and $\ve[y]$, leading to
\begin{align}
\hat{E}_{ij}(s)=&\ \frac{1}{s}\int_{\Omega}d\ve[x]u_i({\bf x})\hat B_j(\ve[x],s) \nonumber\\
& -\frac{1}{s}\int_{S_{\Omega}}dS_l(\ve[x])\kappa_{il}({\bf x})\hat B_j(\ve[x],s)
+\frac{\langle\kappa_{ij}\rangle_0}{s^2},\label{ExpreE}
\end{align}
where we have defined
\begin{align}
	\hat B_j(\ve[x],s)=\int_{\mathbb{R}^d}d\ve[y]\  \hat p(\ve[x],s\vert\ve[y])P_0(\ve[y]) (x_j-y_j).\label{043}
\end{align}
Now comes the key point of our derivation. In order to calculate $B_j$, we first consider another probability distribution $q$ defined as
\begin{align}
	q(\ve[y],t\vert \ve[x])=\frac{p(\ve[x],t \vert\ve[y])P_0(\ve[y])}{P_0(\ve[x])}. \label{DefQ}
\end{align}
Using  Bayes' theorem, we see that $q(\ve[y],t\vert \ve[x])$ is the probability density at the position $\ve[y]$ at time $0$ given that $\ve[x]$ is the position at the later time $t$; $q$ is thus the propagator of the tracer particles under time reversal. The evolution of $q$ satisfies
\begin{align}
\partial_t q(\ve[y],t\vert\ve[x])P_0(\ve[x])=&-\partial_{x_i} [u_i(\ve[x])P_0(\ve[x])q(\ve[y],t\vert\ve[x])]\nonumber\\
&+\partial_{x_i}\partial_{x_j}[\kappa_{ij}(\ve[x])P_0(\ve[x])q(\ve[y],t\vert\ve[x])].
\end{align}
Expanding all the derivatives and using the fact that $HP_0=0$, we find that $q$ satisfies\begin{align}
	\partial_t q(\ve[y],t\vert\ve[x])=&[u_i^*(\ve[x])\partial_{x_i} +\kappa_{ij}(\ve[x])\partial_{x_i}\partial_{x_j}]q(\ve[y],t\vert\ve[x]),\label{BackwardEff}
\end{align}
where
\begin{align}
	u_i^*(\ve[x])&=2\frac{ \partial_{x_j}[\kappa_{ij}(\ve[x])P_0(\ve[x])]}{P_0(\ve[x])}-u_i(\ve[x]), \nonumber\\
					 &= u_i(\ve[x])-2\frac{J_{0i}(\ve[x])}{P_0(\ve[x])}, \label{DefA}
\end{align}
is the drift of the time reversed process and where we have introduced the current in the steady state $J_{i}$ 
\begin{align}
	J_{i}(\ve[x])=u_i(\ve[x])P_0(\ve[x])-\partial_{x_k}[\kappa_{ik}(\ve[x])P_0(\ve[x])].
\end{align} 
If we interpret (\ref{BackwardEff}) as a backward Fokker-Planck equation \cite{gardiner1983handbook}, we deduce that $q$ also satisfies the associated forward Fokker-Planck equation
\begin{align}
	\partial_t & q(\ve[y],t\vert\ve[x])=\nonumber\\
& -\partial_{y_i}[u_i^*(\ve[y]) q(\ve[y],t\vert\ve[x])]+ \partial_{y_i}\partial_{y_j}[\kappa_{ij}(\ve[y])q(\ve[y],t\vert\ve[x])]. \label{FKPq}
\end{align}
Therefore $q$ is  the propagator of fictive particles, moving in an effective drift field $u_i^*$ instead of $u_i$. Let us say a little bit more about the properties of $q$. From its definition (\ref{DefQ}), we immediately see that the initial condition for $q$ is 
\begin{align}
	q(\ve[y],0\vert\ve[x])=\delta(\ve[x]-\ve[y]).
\end{align}
Given that no particles can flow in or out of the obstacles, the time reversed process must also have no-flux boundary conditions at the surface of obstacles. That is to say that
\begin{align}
0=n_i \{u_i^*(\ve[y])&q(\ve[y],t\vert\ve[x])-\partial_{y_j}[\kappa_{ij}(\ve[y])q(\ve[y],t\vert\ve[x])]\}\nonumber\\
=\frac{n_i}{P_0(\ve[x])}\Big(&\{u_i^*(\ve[y])P_0(\ve[y])-\partial_{y_j}[\kappa_{ij}(\ve[y])P_0(\ve[y])]\}p(\ve[x],t\vert \ve[y])\nonumber\\
&-\kappa_{ij}(\ve[y])P_0(\ve[y])\partial_{y_j}p(\ve[x],t\vert \ve[y])\Big),
\end{align}
and as a consequence we see that   $n_i\kappa_{ij}(\ve[y])\partial_{y_j}p(\ve[x],t\vert\ve[y])=0$, which recovers an established boundary condition for the no-flux boundary condition in terms of the starting coordinate ${\bf y}$ \cite{gardiner1983handbook}. 
It is interesting to note that the steady state current of the time reversed process is given by $-J_i$, {\em i.e.} exactly the opposite of the current of the original process. In addition, we see that for currentless steady states,  where $J_i= 0$, we have that  $ u_i^*=u_i$ and thus  the original and time reversed processes are statistically identical in that $p(\ve[x],t\vert \ve[y])= q(\ve[x],t\vert \ve[y])$.
Interestingly, for advection of a particle with constant molecular diffusivity by an incompressible  flow, one can easily show that $P_0= 1/ |\Omega|$ where $|\Omega|$ is the volume of the unit cell and that 
$J_i$ is non-zero and given by $J_i=u_i/|\Omega|$, consequently $u_i^*=-u_i$
,{\em i.e.} the time reversed process has the opposite flow field to the original process.

Eq.~(\ref{043}) may be rewritten in terms of $q$ as
\begin{align}
	\hat B_j(\ve[x],s)=\int_{\mathbb{R}^d}d\ve[y]\  \hat q(\ve[y],s\vert\ve[x])P_0(\ve[x]) (x_j-y_j).\label{043d}
\end{align}
Laplace transforming  the Fokker-Planck equation  (\ref{FKPq}) for $q$ yields  
\begin{align}
s \ & \hat q(\ve[y],s\vert\ve[x])= \delta(\ve[x]-\ve[y]) \nonumber\\
& -\partial_{y_i}[u_i^*(\ve[y]) \hat q(\ve[y],s\vert\ve[x])]+ \partial_{y_i}\partial_{y_j}[\kappa_{ij}(\ve[y])\hat q(\ve[y],s\vert\ve[x])].\label{9049}
\end{align}
Inserting the above equality into Eq.~(\ref{043d}) and integrating by parts over $\ve[y]$, we obtain 
\begin{align}
	\hat B_j(\ve[x],s)=
\frac{1}{s}\int_{\mathbb{R}^d}d\ve[y]\ & P_0(\ve[x])\{-u_j^*(\ve[y]) q(\ve[y],s\vert\ve[x]) \nonumber\\
& +\partial_{y_k}[\kappa_{kj}(\ve[y])\hat q(\ve[y],s\vert\ve[x])] \}.
\end{align}
where we have used the no-flux condition (this time of the time reversed process) at the obstacles boundaries. Integrating by parts the term containing $\kappa_{kj}$, and then switching back to the propagator $p$ instead of $q$ and invoking the periodicity property, we obtain 
\begin{align}
	\hat B_j(\ve[x],s)=
&-\frac{1}{s}\int_{\Omega}d\ve[y]\  P_0(\ve[y]) \hat P(\ve[x],s\vert\ve[y]) u_j^*(\ve[y])  \nonumber\\
 &+ \frac{1}{s}\int_{S_{\Omega}}dS_l(\ve[y]) P_0(\ve[y])\hat P(\ve[x],s\vert\ve[y]) \kappa_{lj}(\ve[y]). \label{942}
\end{align}
Inserting the above expression into Eqs.~(\ref{ExpreE}),(\ref{594}) gives a formal expression for the tensor $\hat\psi_{ij}$. 
Using the fact that $\hat P(\ve[x],s\vert\ve[y])\simeq P_0(\ve[x])/s$ for $s\rightarrow0$, one can check that, for large times, $\psi_{ij}\simeq\mu_i(t)\mu_j(t)$. 
It is now useful  to introduce the difference between the propagator and its stationary value
\begin{equation}
P'({\bf x},t\vert {\bf y}) = P({\bf x},t\vert {\bf y})-P_0({\bf x}),\label{defp'}
\end{equation}
and thus has Laplace transform
\begin{equation}
\hat P'({\bf x},s\vert {\bf y}) = \hat P({\bf x},s\vert {\bf y})-{1\over s}P_0({\bf x}).
\end{equation}
In terms of $\hat P'$, we obtain from Eqs.~(\ref{LinkSigmaPsi}),(\ref{594}),(\ref{ExpreE}), and (\ref{942}) the following final expression for the temporal Laplace transform of the dispersion tensor $\sigma_{ij}(t)$
\begin{align}
&\hat\sigma_{ij}(s)=\frac{2}{s^2}\int_\Omega d{\bf x}\  \kappa_{ij}({\bf x})P_0({\bf x})\nonumber\\
&+\frac{1}{s^2}\int_{\Omega}d\ve[x][u_i(\ve[x])K_j(\ve[x],s)+u_j(\ve[x])K_i(\ve[x],s)] \nonumber\\
&-\frac{1}{s^2}\int_{S_{\Omega}}dS_l(\ve[x])[\kappa_{il}({\bf x})K_j(\ve[x],s)+\kappa_{jl}({\bf x})K_i(\ve[x],s)],\label{SigmaFInal}
\end{align}
with 
\begin{align}
	K_j(\ve[x],s)=
&-\int_{\Omega}d\ve[y]\  P_0(\ve[y]) \hat P'(\ve[x],s\vert\ve[y]) u_j^*(\ve[y])  \nonumber\\
 &+\int_{S_{\Omega}}dS_l(\ve[y]) P_0(\ve[y])\hat P'(\ve[x],s\vert\ve[y]) \kappa_{lj}(\ve[y]). \label{964}
\end{align}
and we recall that $u_i^*$ is the effective drift field after time reversal symmetry defined by Eq.~(\ref{DefA}). Equations (\ref{DefA}),(\ref{SigmaFInal}) and (\ref{964}) give a closed form expression for the Laplace transform of the dispersion tensor $\sigma_{ij}(t)$ at all times and is the key result  of the paper. 

\section{The effective diffusion tensor}\label{eft}
Here we extract the late time effective diffusion tensor $D_{ij}$. The Eqs.  (\ref{SigmaFInal}) and (\ref{964})
are written in Laplace space and the large time limit is thus extracted from the small $s$ behavior, for which 
\begin{align}
	\hat\sigma_{ij}(s)\simeq \frac{2\ D_{ij}}{s^2} \hspace{0.4cm}(s\rightarrow0). 
\end{align}
By definition, 
\begin{align}
	\hat P'(\ve[x],s\vert\ve[y])=\int_0^{\infty}dt \ e^{-s t} [P(\ve[x],t\vert\ve[y])-P_0(\ve[x])].\label{LaplaceTransformP'Def}
\end{align}
Taking $s=0$ in the above expression, we find  that $\hat P'$ is finite at $s=0$ (as 
$P(\ve[x],t\vert\ve[y])\to P_0(\ve[x])$ as $t\to \infty$). Therefore, we can define a function $G$ such that
\begin{align}
	\hat P(\ve[x],0\vert\ve[y]) =\int_0^{\infty}dt \ [P(\ve[x],t\vert\ve[y])-P_0(\ve[x])] \equiv G(\ve[x]\vert\ve[y]). \label{DefG}
\end{align}
Applying the operator $H$ to the above equality, it is clear that $G$ obeys 
\begin{align}
	H_{\ve[x]}G(\ve[x]\vert\ve[y])=\delta(\ve[x]-\ve[y])-P_0(\ve[x]), \label{PseudoGreenFunction}
\end{align}
and also from the conservation of probability $G$ satisfies 
\begin{align}
\int_{\Omega}d\ve[x] \ G(\ve[x]\vert\ve[y])=0. \label{98U05}
\end{align}
The function $G$ is therefore the pseudo-Green's function \cite{barton1989elements} of the operator $H$ (that is, the inverse of $H$ in the subspace orthogonal to the uniform function).

Since $\hat P'(\ve[x],s=0\vert\ve[y])$ is finite, we  see that the inverse Laplace transform at late times can be extracted from Eqs.~(\ref{SigmaFInal}) and (\ref{964}) upon setting $s=0$ in $\hat P'$. This immediately yields 
\begin{align}
D_{ij}=&\int_\Omega d{\bf x}\  \kappa_{ij}({\bf x})P_0({\bf x})\nonumber\\
&+\frac{1}{2}\int_{\Omega}d\ve[x]\ [u_i(\ve[x])f_j(\ve[x])+u_j(\ve[x])f_i(\ve[x])] \nonumber\\
&-\frac{1}{2} \int_{S_{\Omega}}dS_l(\ve[x])[\kappa_{il}({\bf x})f_j(\ve[x])+\kappa_{jl}({\bf x})f_i(\ve[x])], \label{formstat}
\end{align}
with 
\begin{align}
	f_j(\ve[x])=
 &-\int_{\Omega}d\ve[y]\  P_0(\ve[y])G(\ve[x] \vert\ve[y]) u_j^*(\ve[y])  \nonumber\\
 &+\int_{S_{\Omega}}dS_l(\ve[y]) P_0(\ve[y])G(\ve[x]\vert\ve[y]) \kappa_{lj}(\ve[y]). \label{9644545}
\end{align}
For computational purposes, it is useful to find the partial differential equation satisfied by $f_i$. Acting with $H$ on  the above expression and using Eq. (\ref{PseudoGreenFunction}) leads to
\begin{align}
H_{\ve[x]}f_i(\ve[x]) =&  - u_i^*({\bf x})P_0({\bf x}) + P_0(\ve[x]) \int_\Omega d{\bf y} \ 
[u_i^*(\ve[y])P_0({\bf y})]
 \nonumber\\
 &-P_0(\ve[x])\int_{S_{\Omega}}dS_l(\ve[y]) P_0(\ve[y]) \kappa_{li}(\ve[y]),\label{eqfi}
\end{align}
which is valid for $\ve[x]\in\Omega$ (outside the obstacles). 
An equivalent equation for $f_i$ is
\begin{align}
&H_{\ve[x]}f_i(\ve[x]) = 
  - u_i^*({\bf x})P_0({\bf x}) \nonumber\\
&+ P_0(\ve[x]) \int_\Omega d{\bf y} \ 
\{u_i^*(\ve[y])P_0({\bf y}) -  \partial_{y_k}[P_0(\ve[y]) \kappa_{kj}(\ve[y])]\}.\label{eqfi2}
\end{align}
This equation for $f_i(\ve[x])$ (for $\ve[x]$ outside the obstacles) is supplemented by the following conditions. First, because of the conservation of probability, it follows from Eq.~(\ref{98U05}) that
\begin{equation}
\int_\Omega d{\bf x} \ f_i({\bf x})=0. \label{Orthog_f}
\end{equation}
Second,   $f_i(\ve[x])$ is periodic on $\Omega$. Third, at the surface obstacles, $f_i$ satisfies
\begin{align}
	n_j\{u_j({\bf x}) f_i(\ve[x])-\partial_{x_k}[\kappa_{kj}({\bf x})f_i(\ve[x])]\}&=- n_j \kappa_{ji}(\ve[x])P_0(\ve[x]) \label{BC_fi}
\end{align}
It is actually not obvious to derive the boundary conditions (\ref{BC_fi}) from Eq.  (\ref{9644545}). 
The validity of these boundary conditions is checked explicitly in Appendix \ref{AppendixExplicitProof}, where we demonstrate that Eq.~(\ref{9644545}) is the actual solution of the Eqs. (\ref{eqfi2},\ref{Orthog_f},\ref{BC_fi}), proving that our formulation is correct.  

One may alternatively express the diffusion tensor by defining a function $f_i^*$ as 
\begin{align}
	f_i^*(\ve[y])=
 &-\int_{\Omega}d\ve[x]\  P_0(\ve[y])G(\ve[x] \vert\ve[y]) u_i(\ve[x])  \nonumber\\
 &+\int_{S_{\Omega}}dS_l(\ve[x]) P_0(\ve[y])G(\ve[x]\vert\ve[y]) \kappa_{li}(\ve[x]),  
\end{align}
in terms of which the effective diffusion tensor reads 
\begin{align}
D_{ij}=&\int_\Omega d{\bf y}\  \kappa_{ij}({\bf y})P_0({\bf y})\nonumber\\
&+\frac{1}{2}\int_{\Omega}d\ve[y]\ [u_i^*(\ve[y])f_j^*(\ve[y])+u_j^*(\ve[y])f_i^*(\ve[y])] \nonumber\\
&- \frac{1}{2}\int_{S_{\Omega}}dS_l(\ve[y])[\kappa_{il}({\bf y})f_j^*(\ve[y])+\kappa_{jl}({\bf y})f_i^*(\ve[y])].  \label{drf*}
\end{align}
Obviously, 
\begin{align}
	f_i^*(\ve[y])=
 &-\int_{\Omega}d\ve[x]\  P_0(\ve[x])G^*(\ve[y] \vert\ve[x]) u_i(\ve[x])  \nonumber\\
 &+\int_{S_{\Omega}}dS_l(\ve[x]) P_0(\ve[x])G(\ve[y]\vert\ve[x]) \kappa_{li}(\ve[x]),  \label{DeffIStar}
\end{align}
where $G^*$ is the pseudo-Green function associated for the operator $H^*$ of the time reversed process, with the drift field $u^*$, that reads
\begin{align}
H_{\ve[y]}^*(\cdot)=\partial_{y_i}[u_i^*(\ve[y])(\cdot)]-\partial_{y_i}\partial_{y_k}[\kappa_{ij}(\ve[y])(\cdot)]
\end{align}
Hence, $f_i^*$ satisfies the same equations as  $f_i$ if one replaces $u_i$ by $u_i^*$ and vice-versa. As a consequence,  $f_i^*$ is the solution of 
\begin{align}
H_{\ve[y]}^*f_i^*(\ve[y])=&  - u_i({\bf y})P_0({\bf y}) + P_0(\ve[y]) \int_\Omega d{\bf x} \ 
[u_i(\ve[x])P_0({\bf x})]
 \nonumber\\
 &-P_0(\ve[y])\int_{S_{\Omega}}dS_l(\ve[x]) P_0(\ve[x]) \kappa_{li}(\ve[x]) \label{eqgi}
\end{align}
with periodic boundary conditions  and also with the condition
\begin{align}
	n_j\{u_j^*({\bf y}) f_i^*(\ve[y])-\partial_{y_k}[\kappa_{kj}({\bf y})f_i^*(\ve[y])]\}&=-n_j \kappa_{ji}(\ve[y])P_0(\ve[y]) 
\end{align}
at the surface of the obstacles, together with the orthogonality condition
\begin{equation}
	\int_\Omega d{\bf y} \ f_i^*({\bf y})=0.  
\end{equation}
Thus, we have two formulas for the computation of the effective diffusion tensor. 

Having two formulations for the Kubo formula is a useful check on numerical solutions as the two resulting results for the diffusion tensor can be compared. Furthermore, we will now see that both the fields $f_i$ and $f_i^*$ are needed to compute the leading order finite time correction to the  time dependent diffusion tensor.

\section{Late time corrections to the effective diffusion tensor}\label{late}
In section \ref{eft} we have extracted, from the full time  dependent Kubo formula, the effective late time limit of the diffusion tensor $D_{ij}$. While many examples of such formulas exist for special cases, little is known about how the time dependent diffusion tensor relaxes to its asymptotic limit. To our knowledge the only examples known are for diffusion in a periodic potential \cite{dean2014approach} and diffusion in a system with periodic diffusivity \cite{Dean2014PRE}. Both of these examples have steady states with zero current, here we will extend these results to the most general cases both with and without current. 
Here we extract the late time correction tensor $C_{ij}$ that describes the approach to the diffusive limit of the system. 

Expanding Eq.~(\ref{LaplaceTransformP'Def}) in powers of $s$, we obtain
\begin{align} 
\hat P'(\ve[x],s\vert\ve[y])\simeq G(\ve[x]\vert\ve[y])+s\ G^{(2)}(\ve[x]\vert\ve[y])+...\label{ExpansionPPrime}
\end{align}
with
\begin{align}
	G^{(2)}(\ve[x]\vert\ve[y])=-\int_0^{\infty}dt\ t \ [P(\ve[x],t\vert\ve[y])-P_0(\ve[x])]. \label{DefG2}
\end{align}
Since the motion of the tracer particles is Markovian (memoryless), we can write the equality
\begin{align}
P(\ve[x],t\vert\ve[y])=\int_{\Omega} d\ve[z]\  P(\ve[x],t-t_1\vert\ve[z])P(\ve[z],t_1\vert\ve[y]),
\end{align}
which holds for $0<t_1<t$. Integrating this relation over $t_1$ leads to
\begin{align}
t\ P(\ve[x],t\vert\ve[y])=\int_0^t dt_1\int_{\Omega} d\ve[z] \  P(\ve[x],t-t_1\vert\ve[z])P(\ve[z],t_1\vert\ve[y]).
\end{align}
Substracting $tP_0(\ve[x])$  and taking the Laplace transform for small values of $s$, we obtain 
\begin{align}
	G^{(2)}(\ve[x]\vert\ve[y])=- \int_{\Omega} d\ve[z]\  G(\ve[x]\vert\ve[z])G(\ve[z]\vert\ve[y]).\label{G2}
\end{align}
The correction tensor $C_{ij}$ is identified from the relation for small $s$
\begin{align}
\hat\sigma_{ij}(s)\simeq \frac{2}{s^2}( D_{ij}+ s \ C_{ij}+...) \hspace{1cm}( s\rightarrow0).
\end{align}
Inserting this relation into Eqs.~(\ref{SigmaFInal}) and (\ref{964}), and using (\ref{ExpansionPPrime}) and (\ref{G2}) leads to
\begin{align}
&C_{ij}=-\frac{1}{2}\int_\Omega d\ve[z]\Big\{\int_{\Omega}d\ve[x][u_i(\ve[x])f_j(\ve[z])+u_j(\ve[x])f_i(\ve[z])]G(\ve[x]\vert\ve[z]) \nonumber\\
&-\int_{S_{\Omega}}dS_l(\ve[x])[\kappa_{il}({\bf x})f_j(\ve[z])+\kappa_{jl}({\bf x})f_i(\ve[z])]G(\ve[x]\vert\ve[z])\Big\},
\end{align}
where we have used the definition of  $f$ in Eq. (\ref{9644545})
to perform the integration over $\ve[y]$. Using the definition of $f_i^*$ in Eq. (\ref{DeffIStar}) and the relation $G^*(\ve[y]\vert\ve[x])P_0(\ve[x])=G(\ve[x]\vert\ve[y])P_0(\ve[y])$, 
we finally obtain
\begin{align}
C_{ij}=\frac{1}{2}\int_{\Omega}d\ve[z] \ \frac{f_i(\ve[z])f_j^*(\ve[z])+f_j(\ve[z])f_i^*(\ve[z])}{P_0(\ve[z])}.
\end{align}
This relation is a compact and explicit form for the correction tensor $C_{ij}$ and is the main result of this section. One can also derive the above result by decomposing $P'$ in terms of left and right eigenvector of the Fokker-Planck operator $H$. In doing this is is straightforward to see that the temporal corrections at the next order decay as $\exp(-\lambda_1 t)$, where $\lambda_1$ is the lowest positive eigenvalue of $H$ which will be strictly positive given that the domain $\Omega$ is taken to be finite.

\section{Special cases and comparison with existing results}
\label{SectionSpecialCases}
Here we examine the form the Kubo formula takes for a number of systems and compare them to existing results in the literature.

\subsection{Flow in frozen incompressible velocity fields with isotropic constant diffusion tensor}
The Fokker-Planck transport operator  for a tracer advected by an incompressible 
velocity field ${\bf v}$ with constant isotropic diffusivity $\kappa_{ij}({\bf x}) = \kappa_0 \delta_{ij}$ is defined via
\begin{equation}
H_{\ve[x]} = -\kappa_0  \partial_{x_i}  \partial_{x_i} +v_i(\ve[x]) \partial_{x_i} \ .
\end{equation}
Here the drift field is thus $u_i=v_i$. From the incompressibility condition $\partial_{x_i}v_i=0$,
we see that the steady state distribution on the unit cell $\Omega$ is uniform and thus given by
$P_0({\bf x}) = 1/|\Omega|$, where $|\Omega|$ denotes the available volume of the unit cell. The steady state current is thus given by $J_i({\bf x})= v_i({\bf x})/|\Omega|$. We also find that $u_i^*=-u_i$.  Applying Eq. (\ref{formstat}) leads to
\begin{align}
&D_{ij}=\delta_{ij} \kappa_0 +
{1\over 2}\int_\Omega d{\bf x}\ [ f_i({\bf x}) v_j({\bf x})
+ f_j({\bf x}) v_i({\bf x})]\nonumber\\
&-\frac{\kappa_0}{2} \left\{\int_{S_{\Omega}}dS_i(\ve[x]) f_j(\ve[x])+\int_{S_{\Omega}}dS_j(\ve[x]) f_i(\ve[x])\right\}.
\end{align}
The equation satisfied by $f_i$  is a simplified form of Eq. (\ref{eqfi2}):
\begin{equation}
 -\kappa_0 \partial_{x_j}^2f_i +v_j  \partial_{x_j} f_i= {1\over |\Omega|}v_i({\bf x}) - {1\over |\Omega|^2 }\int_\Omega d{\bf y}\  v_i({\bf y}).
\end{equation}
The boundary conditions on the surface of obstacles are  [see Eq.~(\ref{BC_fi})]
\begin{align}
n_j \partial_{x_j}f_i= n_i/\vert\Omega\vert,
\end{align}
(where we have assumed that no fluid enters the obstacles, $v_i n_i=0$). It is then obvious that the last four expressions recover those  in the hydrodynamics-homogenization literature (see \textit{e.g.} Refs.~\cite{carbonell1983dispersion,brenner1980dispersion}), if we identify the function $f_i$ of this paper with $-f_i/\vert\Omega\vert$ to match with the notations of Ref.~\cite{carbonell1983dispersion}.

\subsection{Systems with no steady state current}
A large class of models studied in statistical mechanics, such as diffusion in a periodic 
potential or diffusion in a medium having periodic diffusivity, have no current in the steady state. In these cases  the original and its time-reversed processes have the same Fokker-Planck evolution operator, and consequently the same pseudo-Green's function. However in general we have that
\begin{equation}
G({\bf x}|{\bf y})P_0({\bf y}) = P_0({\bf x}) G^*({\bf y}|{\bf x}),
\end{equation}
where $G^*({\bf y}|{\bf x})$ denotes the pseudo-Green's function for $H^*$. In the case where there is no current we know however that $G^*({\bf y}|{\bf x})=G({\bf x}|{\bf y})$ and thus we find that the operator 
\begin{equation}
K({\bf x},{\bf y})=G({\bf x}|{\bf y})P_0({\bf y})= P_0({\bf x}) G({\bf y}|{\bf x}) 
\end{equation}
is symmetric. In the currentless case we also have
\begin{align}
&D_{ij}= \int_\Omega d{\bf x}\  \kappa_{ij}({\bf x})P_0({\bf x})\nonumber\\
 & -\frac{1}{2}\iint_{\Omega}d\ve[x]d\ve[y][\tilde u_i(\ve[x])\tilde u_j(\ve[y])+\tilde u_j(\ve[x])\tilde u_i(\ve[y])] K({\bf x},{\bf y}), 
\end{align}
from Eq. (\ref{9553}). The operator $K$, being symmetric, must have
real eigenvalues, however $K$ has the same eigenvalues as $G$ which must have
eigenvalues with a positive real part if a steady state regime can be attained. Consequently
under these assumptions $K$ must be a positive operator and we must have that the change in the diagonal terms of the diffusion tensor due to the presence of drift or
inclusions is negative, {\em i.e.}
\begin{equation}
D_{ii}-\langle\kappa_{ii} \rangle_0= -\iint_{\Omega}d\ve[x]d\ve[y] \tilde u_i(\ve[x])\tilde u_i(\ve[y]) K({\bf x},{\bf y})<0,
\end{equation}
that is to say that diffusion in equilibrium systems is slowed down by the presence of the drift. This is physically obvious for say diffusion in periodic potentials where the diffusing particle becomes trapped in local minima of the potential. This slowing down due to drift also occurs for  diffusion in a medium of spatially varying diffusivity where \begin{equation}
H_{\ve[x]} = -  \partial_{x_i} \kappa_{ij}({\bf x}) \partial_{x_j} ,
\end{equation}
here $P_0 = 1/|\Omega|$ and $J_i=0$. The drift field in this case is, in our notation,
given by
\begin{equation}
u_i({\bf x}) = \partial_{x_j}\kappa_{ij}({\bf x}),
\end{equation}
and in terms of the pseudo-Greens' function $G$ we find
\begin{align}
D_{ij}& = {1\over |\Omega|}\int_{\Omega}d\ve[x]\ {\kappa}_{ij}({\bf x}) \nonumber\\
&-{1\over |\Omega|}\iint_{\Omega}d\ve[x]d\ve[y] \ 
[\partial_{x_k}\kappa_{ik}({\bf x})][\partial_{x_l}\kappa_{il}({\bf y})]G({\bf x}|{\bf y}), 
\end{align}
which recovers the formula derived in Ref.~\cite{dean2007effective}.  
We see that the first term in the above is the spatial average  of the local diffusivity tensor while the 
second terms gives a negative contribution when $i=j$. Finally we remark that the case
of diffusion with constant diffusivity $\kappa_0$ in a periodic potential $\Phi$ gives 
$u_i = -\kappa_0\beta\partial_{x_i}\Phi({\bf x})$, where $\beta$ is the inverse temperature.  Here the steady state distribution $P_0$ is given by the Gibbs-Boltzmann distribution on the unit cell $\Omega$
\begin{equation}
P_0({\bf x}) = {\exp\left[-\beta\Phi({\bf x})\right]\over Z_\Omega},
\end{equation}
where $Z_\Omega$ is the partition function normalizing the distribution, thus recovering the
result of Ref.~\cite{dean2007effective} for this particular case.

\section{General results in one dimension}\label{1d}
We consider a generic one dimensional system with periodicity denoted by $L$. Here we can give explicit formula for all terms in the static Kubo formula, although the computation is surprisingly involved. In what follows we will use notation similar to
Refs. \cite{reimann2002diffusion,reimann2001giant} to facilitate comparison with their results. In one dimension the steady state current is constant in space and the  steady state 
probability distribution is given by
\begin{equation}
P_0(x) = JI_+(x),
\end{equation}
where the term $I_+$ reads
\begin{equation}
I_+(x) = {\exp\left(\Gamma(x)\right)\over \kappa(x)}\int_x^\infty dx'\ \exp\left(-\Gamma(x')\right),\label{I+}
\end{equation}
with 
\begin{equation}
\Gamma(x) = \int_0^x dx' {u(x)\over \kappa(x)}.
\end{equation}
Due to the periodicity of $u$ and $\kappa$ the function $\Gamma$ obeys the relation
\begin{equation}
\Gamma(x+L) = \Gamma(x) + \Gamma(L).
\end{equation}
When $\Gamma(L)=0$  the system clearly has a steady state equilibrium distribution with no current. In writing Eq. (\ref{I+}) we have  assumed, without loss of generality, that $\Gamma(L)>0$ so that the integral on the right hand side converges. The steady state current is then obtained from the condition of normalization of $P_0$ and is thus given by
\begin{equation}
J = {1\over \int_0^L dx\  I_+(x)},\label{jo1d}
\end{equation} 
and consequently the effective drift is given by $V =JL$.

To compute the effective diffusion constant we use the representation for $D$ given 
in Eq. (\ref{drf*}) in its one dimensional version, that is to say
\begin{equation}
D = \int_0^Ldx\  \kappa(x)P_0(x)\nonumber\\
+\int_0^L dx f^*(x)u^*(x).\label{1dd1}
\end{equation}
We now define $f^*(x) = g(x)P_0(x)$ and write the diffusion constant as
\begin{equation}
D = \int_0^Ldx\  \kappa(x)P_0(x)\nonumber\\
+\int_0^L dx g(x)\left\{J - {d\over dx}[\kappa(x)P_0(x)]\right\}.\label{1dd1}
\end{equation}
Defining $g(x) = -x + h(x)$  we find that $h$ obeys
\begin{equation}
\kappa {d^2h\over dx^2} + u {dh\over dx} = JL.
\end{equation}
A first integration of this equation for $h$ yields
\begin{equation}
{dh(x)\over dx} = JL I_-(x), 
\end{equation}
where
\begin{equation}
I_-(x)= {\exp\left(-\Gamma(x)\right)}\int_{-\infty}^x dx'\ {\exp\left(\Gamma(x')\right)\over \kappa(x')},\label{I-}
\end{equation}
and where we have used the fact that $dg/dx = -1 + dh/dx$ is periodic and thus $dh/dx$ must be periodic.  Integrating again yields
\begin{equation}
h(x) = J_0L \int_0^x dx'\  I_-(x') + C \label{eqh}
\end{equation}
where $C$ is an integration constant that will be determined from the orthogonality condition $\int_0^L dx\ P_0(x) g(x)=0$. It is not immediately obvious from Eq. (\ref{eqh}) that $g(x) = -x + h$ is periodic, however it is easy to see that
\begin{equation}
g(x+L) = g(x) - L + J L \int_0^L dx\ I_-(x) .\label{stepg}
\end{equation}
Now by integration by parts one finds
\begin{equation}
\int_0^L dx\ I_-(x)  = \int_0^L dx\ I_+(x),
\end{equation}
and using this and Eq. (\ref{jo1d}) in Eq. (\ref{stepg}) yields $g(x)=g(x+L)$. Incidentally this shows that we may also write 
\begin{equation}
J = {1\over \int_0^L dx\  I_-(x)}.
\end{equation}
To proceed it is useful to define the two new functions
\begin{eqnarray}
A_+(x) &=& \int_{-\infty}^x dx'\ {\exp\left(\Gamma(x')\right)\over \kappa(x')},\\
A_-(x) &=& \int_x^{\infty} dx'\ {\exp\left(-\Gamma(x')\right)}.
\end{eqnarray}
Further more one can show that
\begin{eqnarray}
A_+(x+L) &=& \exp\left(\Gamma(L)\right)A_+(x)\label{h+1},\\
A_-(x+L) &=&  \exp\left(-\Gamma(L)\right)A_-(x)\label{h-1}.
\end{eqnarray}
In terms of these two functions we can write  $I_+$ and $I_-$ as
\begin{eqnarray}
I_+(x) &=& {dA_+(x)\over dx} A_-(x)\label{h+I},\\
I_-(x) &=& -{dA_-(x)\over dx} A_+(x).\label{h-I}
\end{eqnarray}
Computing the constant $C$ in Eq.~(\ref{eqh}) using the orthogonality relation then yields
\begin{align}
&D = \int _0^L dx \Big[ \kappa(x) P_0(x) {dh\over dx} \nonumber\\
&+J\left[-x + h(x) + LxP_0(x)
-Lh(x)P_0(x)\right]\Big], \label{stepde}
\end{align}
where we have carried out an integration by parts in the last term of Eq. (\ref{1dd1}).
The function $h$ is explicitly given by
\begin{equation}
h(x) = JL \int_0^x dx'\  I_-(x'),
\end{equation}
and using the expression Eq. (\ref{h-I}) for $I_-$ it can be written as
\begin{align}
h(x) = -JL\Bigg[ A_+(x)A_-(x) - A_+(0)A_-(0)&\nonumber\\
 -\int_0^x dx'\ {P_0(x')\over J}&\Bigg].
\end{align}
We also, by integration by parts, note the identity
\begin{equation}
\int_0^L dx\ \int_0^xdx'\ P_0(x') = L - \int_0^L dx \ xP_0(x).
\end{equation}
Using these relations  in Eq. (\ref{stepde}), after some algebra, we obtain the compact expression
\begin{equation}
D = {L^2\int_0^L dx\ \kappa(x) I_\pm(x)^2 I_\mp (x)  \over \int_0^L dx\  I_\pm(x)^3}, \label{d1dfinal}
\end{equation} 
where $\pm$ indicates that the index can be taken to be $+$ or $-$. One can verify that 
Eq. (\ref{d1dfinal}) agrees with the results of Reimann \textit{et al.} \cite{reimann2002diffusion,reimann2001giant} for the case of diffusion in a tilted  potential $\phi(x) = V(x) -fx$, where $V$ is periodic. It also agrees with the formulas presented in Ref. \cite{reguera2006entropic} where the local mobility is also varying. The result given here shows how 
the diffusion constant of any one dimensional system with periodic diffusivity and drift can be obtained.

\section{Conclusions} 

The problem of computing effective transport coefficients occurs in many settings of fundamental physics and has many applications. Many results exist in the literature, for example in fluid mechanics using homogenization theory 
\cite{rosenbluth1987effective,Shraiman1987,McCarty1988,majda1999simplified,brenner1980dispersion,rubinstein1986dispersion,quintard1994convection,alshare2010modeling,souto1997dispersion,souto1997dispersion1,quintard1993transport}
as well as in statistical physics \cite{dean2007effective,derrida1983velocity,zwanzig1988diffusion,deGennes1975brownian,lifson1962self} where both exact and approximate results exist for problems such as
diffusion in periodic or random potentials and diffusion in media with locally varying diffusivity. In these latter studies most results have been obtained for systems with equilibrium (currentless) steady states, given for example by the Gibbs Boltzmann distribution for diffusion in a periodic potential. 

More recently results have been derived for diffusion
in systems with non zero currents \cite{reimann2001giant,lindner2001optimal,reimann2002diffusion,reimann2008weak,lindner2002,reguera2006entropic,burada2008entropic}
In such non equilibrium systems interesting new physics such as massively increased diffusivity due to an external applied  force has been 
discovered. This enhancement occurs, in the weak noise limit, at the external field where all
minima disappear from the overall potential to be replaced by points of inflection \cite{reimann2001giant}. The difference between different trajectories at this critical point is enhanced  and dispersion in thus increased. Using the Kubo formula, given here, one could investigate the effect of an externally applied field on diffusion in higher dimensional periodic potentials. The formula
given here cannot in general be evaluated analytically in higher than one dimension, however their numerical evaluation is straightforward and the transport coefficients can be accurately determined using standard packages to solve partial differential equations. Of course many of the systems considered here can be studied via numerical simulations 
where one integrates the corresponding SDE, however the simulation approach suffers from the need to estimate the errors due to statistical fluctuations and in the case of 
obstacles the correct imposition of the no-flux boundary condition requires careful treatment and is far from being obvious \cite{Peters2002,Barenbrug2002,lamm1983extended}. 

Many other problems can be tackled using the present approach, in Ref.~\cite{guerin2015} it was shown how the presence of an externally applied uniform field on a system with spatially varying diffusivity modifies the dispersion of a cloud of tracer particles. It was shown that the diffusion in the direction of the applied force could be hugely increased for systems in two or more dimensions. In addition it was shown that the components of the effective diffusion tensor can exhibit counter intuitive 
monotonic behavior. The results thus imply that one can control the dispersion of a diffusing cloud with an external field, and such an effect may well have useful applications. 

The formula given here are extremely general and should be valuable for the interpretation of 
experiments. For instance, potential fields in which colloidal particles diffuse are often generated by laser \cite{DalleFerrier2011dynamics,Evstigneev2008,evers2013colloids,hanes2012colloids}, in which case absorption of laser light can also lead to temperature gradients. The variation 
of the local temperature will clearly influence local transport properties
through the temperature dependence of the surrounding fluid 
viscosity as well as due to thermodynamic forces associated with temperature gradients, {\em i.e}
the Soret effect \cite{wurger2010thermal}.
 
Finally there has been much recent study of fluctuation dissipation relations (FDR) in 
non-equilibrium systems \cite{speck2006restoring,blickle2007einstein,baiesi2011modified,maes2013fluctuation}. A notable example of this type of FDR is the Stokes-Einstein
relation between the differential response of the mean velocity $V_i$  and the diffusion
tensor. In our preliminary study \cite{guerin2015}, we showed how the presence of a steady state current leads to corrections to the usual Stokes-Einstein formula. However the general analysis of the 
correction term needs further study in order to fully understand the non-equilibrium physics
it encodes, as well as to make contact with the literature on the subject \cite{speck2006restoring,blickle2007einstein,baiesi2011modified,maes2013fluctuation}. In the literature in question results are most often given in terms of an integral over a time dependent violation factor, the results given here could be useful to understand these results in terms of the static, time independent quantities, used here.

\appendix

\section{Details on the numerical simulations}
\label{AppSimu}
Here we briefly describe the algorithm used to simulate the motion of a Brownian walker in a 2D array of circular disks, in the presence of a force $F$, leading to the results presented in Fig.\ref{figForceObstacles}(b). Let $\ve[X]$ be the position at time $t$ of the walker, then we define $\ve[e]_{\alpha}$ the unit vector with direction $\ve[X]-\ve[X]_c$, with $\ve[X]_c$ the nearest disk center. $\ve[e]_{\alpha}$ makes an angle $\alpha$ with the $x-$direction. During the time step $\Delta t$, the motion in the direction $\ve[e]_{\alpha}$ is modified by the increment 
\begin{align}
dX_{\perp}=\sqrt{\Delta t}f_1(r_{\perp}/\sqrt{\Delta t})+u_{\parallel} f_2(r_{\perp}/\sqrt{\Delta t})+h\cos\alpha\Delta t \label{IncrPar}
\end{align}
Here $h=\beta F D$, $r_{\perp}$ is the distance to the nearest disk surface, and $f_1$ and $f_2$ are the functions calculated by Peters \textit{et al.} in the case of reflecting boundaries \cite{Peters2002}, and $u_{\parallel}=\pm  \sqrt{D\Delta t} $ with probabilities $1/2$. In the direction parallel to the nearest surface obstacles, the increment is $dX_{\parallel}=-h\sin\alpha\Delta t+u$,  where $u$ is a stochastic variable with zero mean and variance $\langle u^2\rangle=2 D \Delta t$.  With this algorithm, when the particle is close to the boundaries, $r_{\perp}\sim\sqrt{\Delta t}$, the terms in $f_1$ and $f_2$ dominate the increment in the perpendicular direction in Eq. (\ref{IncrPar}) if $\Delta t$ is small enough. These terms account for the fact that, since trajectories that cross the surface of the obstacles are forbidden, the averaging over authorized trajectories between $t$ and $t+\Delta t$ gives rise to an effective drift oriented towards the exterior of the obstacles, described by the term $f_1$, and also modifies the variance of these trajectories (term $f_2$), see Ref  \cite{Peters2002} for details. If the particle is far from the obstacles, $r_{\perp}\gg \sqrt{\Delta t}$, then $f_1\rightarrow0$ and $f_2\rightarrow\sqrt{2}$ and one recovers Brownian motion under force in the absence of obstacles. Although the algorithm of Ref \cite{Peters2002} was proposed only for one-dimensional geometries, we expect that it is also valid in the present 2D case, since for sufficiently small $\Delta t$ the curvature of the surface should play no role. For the results of Fig.~\ref{figForceObstacles}(b), the time step was $\Delta t=10^{-5}L^2/D$ and was checked to be sufficiently small so that convergence was reached for the calculation of $D_{ij}$ which were estimated over $500,000$ trajectories. The results of simulations match with the predictions of our (exact) Kubo formulas.

\section{Boundary condition (\ref{BC_fi}) for the function $f_i$}
\label{AppendixExplicitProof}
In this appendix, we show that the formulation of the problem under the form of Eqs.~(\ref{eqfi2},\ref{Orthog_f},\ref{BC_fi}) is correct by checking that Eq.~(\ref{9644545}) is the actual solution of these equations. We consider the quantity
\begin{align}
M_i(\ve[x])\equiv\int_{\Omega}d\ve[y] \left\{ [H_{\ve[y]}^{\dagger}G(\ve[x]\vert\ve[y])] f_i(\ve[y]) -G(\ve[x]\vert\ve[y]) [H_{\ve[y]}f_i(\ve[y])] \right\} \label{059304}
\end{align}
where the adjoint operator reads
\begin{align}
	H_{\ve[y]}^{\dagger}=-u_i(\ve[y])\partial_{y_i}-\kappa_{ij}(\ve[y])\partial_{y_i}\partial_{y_j}.
\end{align}
The pseudo-Green function $G$ can be shown to satisfy the adjoint equation
\begin{align}
	H_{\ve[y]}^{\dagger}G(\ve[x]\vert\ve[y])=\delta(\ve[x]-\ve[y])-P_0(\ve[x]), \label{PseudoGreenFunctionAdjoint}
\end{align}
The following condition also holds
\begin{align}
\int_{\Omega} d\ve[y]\ P_0(\ve[y]) G(\ve[x]\vert\ve[y])=0. \label{OrthogGDagger}
\end{align}
Using the above relations and Eqs.~(\ref{eqfi}) and (\ref{Orthog_f}) we deduce that
\begin{align}
	M_i(\ve[x])=f_i(\ve[x])+ \int_{\Omega}d\ve[y]\  P_0(\ve[y])G(\ve[x] \vert\ve[y]) u_i^*(\ve[y]). \label{754}
\end{align}
On the other hand, using the explicit expressions of $H$  [Eq.(\ref{fp})] and  integrating by parts two times the term containing $H$ in Eq. (\ref{059304}), we obtain
\begin{align} 
&	M_i(\ve[x])=\nonumber\\
&\int_{S_{\Omega}}dS_j(\ve[y]) G(\ve[x]\vert\ve[y])\{-u_j(\ve[y])f_i(\ve[y])+\partial_{y_k}[\kappa_{jk}(\ve[y])f_i(\ve[y])]\}\nonumber\\
&-\int_{S_{\Omega}}dS_j(\ve[y])f(\ve[y])\kappa_{kj}(\ve[y])\partial_{y_k}G(\ve[x]\vert\ve[y]).
\end{align}
The last term of this equation vanishes, since the relation $dS_j\kappa_{kj}(\ve[y])P(\ve[x],t\vert\ve[y])=0$ is known for reflecting boundaries \cite{gardiner1983handbook}. Then, using the boundary conditions (\ref{BC_fi}), we obtain
\begin{align}
M_i(\ve[x])=&\int_{S_{\Omega}}dS_j(\ve[y]) G(\ve[x]\vert\ve[y])\kappa_{jk}(\ve[x]),
\end{align}
which, when compared to Eq.~(\ref{754}), is exactly Eq.~(\ref{9644545}), thereby proving that the formulation of Eqs.~(\ref{eqfi2},\ref{BC_fi},\ref{Orthog_f}) is correct.


\begin{thebibliography}{50}
\expandafter\ifx\csname natexlab\endcsname\relax\def\natexlab#1{#1}\fi
\expandafter\ifx\csname bibnamefont\endcsname\relax
  \def\bibnamefont#1{#1}\fi
\expandafter\ifx\csname bibfnamefont\endcsname\relax
  \def\bibfnamefont#1{#1}\fi
\expandafter\ifx\csname citenamefont\endcsname\relax
  \def\citenamefont#1{#1}\fi
\expandafter\ifx\csname url\endcsname\relax
  \def\url#1{\texttt{#1}}\fi
\expandafter\ifx\csname urlprefix\endcsname\relax\def\urlprefix{URL }\fi
\providecommand{\bibinfo}[2]{#2}
\providecommand{\eprint}[2][]{\url{#2}}

\bibitem[{\citenamefont{Van~Kampen}(2007)}]{VanKampen1992}
\bibinfo{author}{\bibfnamefont{N.}~\bibnamefont{Van~Kampen}},
  \emph{\bibinfo{title}{Stochastic Processes in Physics and Chemistry, Third
  Edition}} (\bibinfo{publisher}{North-Holland, Amsterdam},
  \bibinfo{year}{2007}).

\bibitem[{\citenamefont{{\O}ksendal}(2003)}]{oksendal2003stochastic}
\bibinfo{author}{\bibfnamefont{B.}~\bibnamefont{{\O}ksendal}},
  \emph{\bibinfo{title}{Stochastic differential equations}}
  (\bibinfo{publisher}{Springer, New-York}, \bibinfo{year}{2003}).

\bibitem[{\citenamefont{Gardiner}(1985)}]{gardiner1983handbook}
\bibinfo{author}{\bibfnamefont{C.}~\bibnamefont{Gardiner}},
  \emph{\bibinfo{title}{Handbook of stochastic methods for physics, chemistry
  and the natural sciences, second edition}} (\bibinfo{year}{1985}).

\bibitem[{\citenamefont{Brenner}(1961)}]{brenner1961slow}
\bibinfo{author}{\bibfnamefont{H.}~\bibnamefont{Brenner}},
  \bibinfo{journal}{Chem. Eng. Sci.} \textbf{\bibinfo{volume}{16}},
  \bibinfo{pages}{242} (\bibinfo{year}{1961}).

\bibitem[{\citenamefont{W{\"u}rger}(2010)}]{wurger2010thermal}
\bibinfo{author}{\bibfnamefont{A.}~\bibnamefont{W{\"u}rger}},
  \bibinfo{journal}{Rep. Progr. Phys.} \textbf{\bibinfo{volume}{73}},
  \bibinfo{pages}{126601} (\bibinfo{year}{2010}).

\bibitem[{\citenamefont{Daumas et~al.}(2003)\citenamefont{Daumas, Destainville,
  Millot, Lopez, Dean, and Salom{\'e}}}]{daumas2003confined}
\bibinfo{author}{\bibfnamefont{F.}~\bibnamefont{Daumas}},
  \bibinfo{author}{\bibfnamefont{N.}~\bibnamefont{Destainville}},
  \bibinfo{author}{\bibfnamefont{C.}~\bibnamefont{Millot}},
  \bibinfo{author}{\bibfnamefont{A.}~\bibnamefont{Lopez}},
  \bibinfo{author}{\bibfnamefont{D.}~\bibnamefont{Dean}}, \bibnamefont{and}
  \bibinfo{author}{\bibfnamefont{L.}~\bibnamefont{Salom{\'e}}},
  \bibinfo{journal}{Biophys. J.} \textbf{\bibinfo{volume}{84}},
  \bibinfo{pages}{356} (\bibinfo{year}{2003}).

\bibitem[{\citenamefont{Reits and Neefjes}(2001)}]{reits2001fixed}
\bibinfo{author}{\bibfnamefont{E.~A.} \bibnamefont{Reits}} \bibnamefont{and}
  \bibinfo{author}{\bibfnamefont{J.~J.} \bibnamefont{Neefjes}},
  \bibinfo{journal}{Nat. Cell Biol.} \textbf{\bibinfo{volume}{3}},
  \bibinfo{pages}{E145} (\bibinfo{year}{2001}).

\bibitem[{\citenamefont{Condamin et~al.}(2007)\citenamefont{Condamin,
  B{\'e}nichou, Tejedor, Voituriez, and Klafter}}]{condamin2007}
\bibinfo{author}{\bibfnamefont{S.}~\bibnamefont{Condamin}},
  \bibinfo{author}{\bibfnamefont{O.}~\bibnamefont{B{\'e}nichou}},
  \bibinfo{author}{\bibfnamefont{V.}~\bibnamefont{Tejedor}},
  \bibinfo{author}{\bibfnamefont{R.}~\bibnamefont{Voituriez}},
  \bibnamefont{and} \bibinfo{author}{\bibfnamefont{J.}~\bibnamefont{Klafter}},
  \bibinfo{journal}{Nature} \textbf{\bibinfo{volume}{450}}, \bibinfo{pages}{77}
  (\bibinfo{year}{2007}).

\bibitem[{\citenamefont{Taylor}(1953)}]{taylor1953dispersion}
\bibinfo{author}{\bibfnamefont{G.}~\bibnamefont{Taylor}},
  \bibinfo{journal}{Proc. R. Soc. Lon. A} \textbf{\bibinfo{volume}{219}},
  \bibinfo{pages}{186} (\bibinfo{year}{1953}).

\bibitem[{\citenamefont{Rosenbluth et~al.}(1987)\citenamefont{Rosenbluth, Berk,
  Doxas, and Horton}}]{rosenbluth1987effective}
\bibinfo{author}{\bibfnamefont{M.}~\bibnamefont{Rosenbluth}},
  \bibinfo{author}{\bibfnamefont{H.}~\bibnamefont{Berk}},
  \bibinfo{author}{\bibfnamefont{I.}~\bibnamefont{Doxas}}, \bibnamefont{and}
  \bibinfo{author}{\bibfnamefont{W.}~\bibnamefont{Horton}},
  \bibinfo{journal}{Phys. Fluids} \textbf{\bibinfo{volume}{30}},
  \bibinfo{pages}{2636} (\bibinfo{year}{1987}).

\bibitem[{\citenamefont{Shraiman}(1987)}]{Shraiman1987}
\bibinfo{author}{\bibfnamefont{B.~I.} \bibnamefont{Shraiman}},
  \bibinfo{journal}{Phys. Rev. A} \textbf{\bibinfo{volume}{36}},
  \bibinfo{pages}{261} (\bibinfo{year}{1987}).

\bibitem[{\citenamefont{McCarty and Horsthemke}(1988)}]{McCarty1988}
\bibinfo{author}{\bibfnamefont{P.}~\bibnamefont{McCarty}} \bibnamefont{and}
  \bibinfo{author}{\bibfnamefont{W.}~\bibnamefont{Horsthemke}},
  \bibinfo{journal}{Phys. Rev. A} \textbf{\bibinfo{volume}{37}},
  \bibinfo{pages}{2112} (\bibinfo{year}{1988}).

\bibitem[{\citenamefont{Majda and Kramer}(1999)}]{majda1999simplified}
\bibinfo{author}{\bibfnamefont{A.~J.} \bibnamefont{Majda}} \bibnamefont{and}
  \bibinfo{author}{\bibfnamefont{P.~R.} \bibnamefont{Kramer}},
  \bibinfo{journal}{Phys. Rep.} \textbf{\bibinfo{volume}{314}},
  \bibinfo{pages}{237} (\bibinfo{year}{1999}).

\bibitem[{\citenamefont{Brenner}(1980)}]{brenner1980dispersion}
\bibinfo{author}{\bibfnamefont{H.}~\bibnamefont{Brenner}},
  \bibinfo{journal}{Philos. Tr. R. Soc. A} \textbf{\bibinfo{volume}{297}},
  \bibinfo{pages}{81} (\bibinfo{year}{1980}).

\bibitem[{\citenamefont{Rubinstein and Mauri}(1986)}]{rubinstein1986dispersion}
\bibinfo{author}{\bibfnamefont{J.}~\bibnamefont{Rubinstein}} \bibnamefont{and}
  \bibinfo{author}{\bibfnamefont{R.}~\bibnamefont{Mauri}},
  \bibinfo{journal}{SIAM J. Appl. Math.} \textbf{\bibinfo{volume}{46}},
  \bibinfo{pages}{1018} (\bibinfo{year}{1986}).

\bibitem[{\citenamefont{Quintard and Whitaker}(1994)}]{quintard1994convection}
\bibinfo{author}{\bibfnamefont{M.}~\bibnamefont{Quintard}} \bibnamefont{and}
  \bibinfo{author}{\bibfnamefont{S.}~\bibnamefont{Whitaker}},
  \bibinfo{journal}{Adv. Water. Ress.} \textbf{\bibinfo{volume}{17}},
  \bibinfo{pages}{221} (\bibinfo{year}{1994}).

\bibitem[{\citenamefont{Alshare et~al.}(2010)\citenamefont{Alshare, Strykowski,
  and Simon}}]{alshare2010modeling}
\bibinfo{author}{\bibfnamefont{A.}~\bibnamefont{Alshare}},
  \bibinfo{author}{\bibfnamefont{P.}~\bibnamefont{Strykowski}},
  \bibnamefont{and} \bibinfo{author}{\bibfnamefont{T.}~\bibnamefont{Simon}},
  \bibinfo{journal}{Int. J. Heat Mass Transfer} \textbf{\bibinfo{volume}{53}},
  \bibinfo{pages}{2294} (\bibinfo{year}{2010}).

\bibitem[{\citenamefont{Souto and
  Moyne}(1997{\natexlab{a}})}]{souto1997dispersion}
\bibinfo{author}{\bibfnamefont{H.~P.~A.} \bibnamefont{Souto}} \bibnamefont{and}
  \bibinfo{author}{\bibfnamefont{C.}~\bibnamefont{Moyne}},
  \bibinfo{journal}{Phys. Fluids} \textbf{\bibinfo{volume}{9}},
  \bibinfo{pages}{2253} (\bibinfo{year}{1997}{\natexlab{a}}).

\bibitem[{\citenamefont{Souto and
  Moyne}(1997{\natexlab{b}})}]{souto1997dispersion1}
\bibinfo{author}{\bibfnamefont{H.~P.~A.} \bibnamefont{Souto}} \bibnamefont{and}
  \bibinfo{author}{\bibfnamefont{C.}~\bibnamefont{Moyne}},
  \bibinfo{journal}{Phys. Fluids} \textbf{\bibinfo{volume}{9}},
  \bibinfo{pages}{2243} (\bibinfo{year}{1997}{\natexlab{b}}).

\bibitem[{\citenamefont{Quintard and Whitaker}(1993)}]{quintard1993transport}
\bibinfo{author}{\bibfnamefont{M.}~\bibnamefont{Quintard}} \bibnamefont{and}
  \bibinfo{author}{\bibfnamefont{S.}~\bibnamefont{Whitaker}},
  \bibinfo{journal}{Chem. Eng. Sci.} \textbf{\bibinfo{volume}{48}},
  \bibinfo{pages}{2537} (\bibinfo{year}{1993}).

\bibitem[{\citenamefont{Dean et~al.}(2007)\citenamefont{Dean, Drummond, and
  Horgan}}]{dean2007effective}
\bibinfo{author}{\bibfnamefont{D.~S.} \bibnamefont{Dean}},
  \bibinfo{author}{\bibfnamefont{I.}~\bibnamefont{Drummond}}, \bibnamefont{and}
  \bibinfo{author}{\bibfnamefont{R.}~\bibnamefont{Horgan}},
  \bibinfo{journal}{J. Stat. Mech.} \textbf{\bibinfo{volume}{2007}},
  \bibinfo{pages}{P07013} (\bibinfo{year}{2007}).

\bibitem[{\citenamefont{Derrida}(1983)}]{derrida1983velocity}
\bibinfo{author}{\bibfnamefont{B.}~\bibnamefont{Derrida}}, \bibinfo{journal}{J.
  Stat. Phys.} \textbf{\bibinfo{volume}{31}}, \bibinfo{pages}{433}
  (\bibinfo{year}{1983}).

\bibitem[{\citenamefont{Zwanzig}(1988)}]{zwanzig1988diffusion}
\bibinfo{author}{\bibfnamefont{R.}~\bibnamefont{Zwanzig}},
  \bibinfo{journal}{Proc. Natl. Acad. Sci. U S A}
  \textbf{\bibinfo{volume}{85}}, \bibinfo{pages}{2029} (\bibinfo{year}{1988}).

\bibitem[{\citenamefont{De~Gennes}(1975)}]{deGennes1975brownian}
\bibinfo{author}{\bibfnamefont{P.}~\bibnamefont{De~Gennes}},
  \bibinfo{journal}{J. Stat. Phys.} \textbf{\bibinfo{volume}{12}},
  \bibinfo{pages}{463} (\bibinfo{year}{1975}).

\bibitem[{\citenamefont{Lifson and Jackson}(1962)}]{lifson1962self}
\bibinfo{author}{\bibfnamefont{S.}~\bibnamefont{Lifson}} \bibnamefont{and}
  \bibinfo{author}{\bibfnamefont{J.~L.} \bibnamefont{Jackson}},
  \bibinfo{journal}{J. Chem. Phys.} \textbf{\bibinfo{volume}{36}},
  \bibinfo{pages}{2410} (\bibinfo{year}{1962}).

\bibitem[{\citenamefont{Reimann et~al.}(2001)\citenamefont{Reimann, Van~den
  Broeck, Linke, H{\"a}nggi, Rubi, and P{\'e}rez-Madrid}}]{reimann2001giant}
\bibinfo{author}{\bibfnamefont{P.}~\bibnamefont{Reimann}},
  \bibinfo{author}{\bibfnamefont{C.}~\bibnamefont{Van~den Broeck}},
  \bibinfo{author}{\bibfnamefont{H.}~\bibnamefont{Linke}},
  \bibinfo{author}{\bibfnamefont{P.}~\bibnamefont{H{\"a}nggi}},
  \bibinfo{author}{\bibfnamefont{J.}~\bibnamefont{Rubi}}, \bibnamefont{and}
  \bibinfo{author}{\bibfnamefont{A.}~\bibnamefont{P{\'e}rez-Madrid}},
  \bibinfo{journal}{Phys. Rev. Lett.} \textbf{\bibinfo{volume}{87}},
  \bibinfo{pages}{010602} (\bibinfo{year}{2001}).

\bibitem[{\citenamefont{Lindner et~al.}(2001)\citenamefont{Lindner, Kostur, and
  Schimansky-Geier}}]{lindner2001optimal}
\bibinfo{author}{\bibfnamefont{B.}~\bibnamefont{Lindner}},
  \bibinfo{author}{\bibfnamefont{M.}~\bibnamefont{Kostur}}, \bibnamefont{and}
  \bibinfo{author}{\bibfnamefont{L.}~\bibnamefont{Schimansky-Geier}},
  \bibinfo{journal}{Fluct. Noise Lett.} \textbf{\bibinfo{volume}{01}},
  \bibinfo{pages}{R25} (\bibinfo{year}{2001}).

\bibitem[{\citenamefont{Reimann et~al.}(2002)\citenamefont{Reimann, Van~den
  Broeck, Linke, H{\"a}nggi, Rubi, and
  P{\'e}rez-Madrid}}]{reimann2002diffusion}
\bibinfo{author}{\bibfnamefont{P.}~\bibnamefont{Reimann}},
  \bibinfo{author}{\bibfnamefont{C.}~\bibnamefont{Van~den Broeck}},
  \bibinfo{author}{\bibfnamefont{H.}~\bibnamefont{Linke}},
  \bibinfo{author}{\bibfnamefont{P.}~\bibnamefont{H{\"a}nggi}},
  \bibinfo{author}{\bibfnamefont{J.}~\bibnamefont{Rubi}}, \bibnamefont{and}
  \bibinfo{author}{\bibfnamefont{A.}~\bibnamefont{P{\'e}rez-Madrid}},
  \bibinfo{journal}{Phys. Rev. E} \textbf{\bibinfo{volume}{65}},
  \bibinfo{pages}{031104} (\bibinfo{year}{2002}).

\bibitem[{\citenamefont{Reimann and Eichhorn}(2008)}]{reimann2008weak}
\bibinfo{author}{\bibfnamefont{P.}~\bibnamefont{Reimann}} \bibnamefont{and}
  \bibinfo{author}{\bibfnamefont{R.}~\bibnamefont{Eichhorn}},
  \bibinfo{journal}{Phys. Rev. Lett.} \textbf{\bibinfo{volume}{101}},
  \bibinfo{pages}{180601} (\bibinfo{year}{2008}).

\bibitem[{\citenamefont{Lindner and Schimansky-Geier}(2002)}]{lindner2002}
\bibinfo{author}{\bibfnamefont{B.}~\bibnamefont{Lindner}} \bibnamefont{and}
  \bibinfo{author}{\bibfnamefont{L.}~\bibnamefont{Schimansky-Geier}},
  \bibinfo{journal}{Phys. Rev. Lett.} \textbf{\bibinfo{volume}{89}},
  \bibinfo{pages}{230602} (\bibinfo{year}{2002}).

\bibitem[{\citenamefont{Reguera et~al.}(2006)\citenamefont{Reguera, Schmid,
  Burada, Rubi, Reimann, and H{\"a}nggi}}]{reguera2006entropic}
\bibinfo{author}{\bibfnamefont{D.}~\bibnamefont{Reguera}},
  \bibinfo{author}{\bibfnamefont{G.}~\bibnamefont{Schmid}},
  \bibinfo{author}{\bibfnamefont{P.~S.} \bibnamefont{Burada}},
  \bibinfo{author}{\bibfnamefont{J.}~\bibnamefont{Rubi}},
  \bibinfo{author}{\bibfnamefont{P.}~\bibnamefont{Reimann}}, \bibnamefont{and}
  \bibinfo{author}{\bibfnamefont{P.}~\bibnamefont{H{\"a}nggi}},
  \bibinfo{journal}{Phys. Rev. Lett.} \textbf{\bibinfo{volume}{96}},
  \bibinfo{pages}{130603} (\bibinfo{year}{2006}).

\bibitem[{\citenamefont{Burada et~al.}(2008)\citenamefont{Burada, Schmid,
  Talkner, H{\"a}nggi, Reguera, and Rubi}}]{burada2008entropic}
\bibinfo{author}{\bibfnamefont{P.}~\bibnamefont{Burada}},
  \bibinfo{author}{\bibfnamefont{G.}~\bibnamefont{Schmid}},
  \bibinfo{author}{\bibfnamefont{P.}~\bibnamefont{Talkner}},
  \bibinfo{author}{\bibfnamefont{P.}~\bibnamefont{H{\"a}nggi}},
  \bibinfo{author}{\bibfnamefont{D.}~\bibnamefont{Reguera}}, \bibnamefont{and}
  \bibinfo{author}{\bibfnamefont{J.}~\bibnamefont{Rubi}},
  \bibinfo{journal}{BioSystems} \textbf{\bibinfo{volume}{93}},
  \bibinfo{pages}{16} (\bibinfo{year}{2008}).

\bibitem[{\citenamefont{Gu\'erin and Dean}(2015)}]{guerin2015}
\bibinfo{author}{\bibfnamefont{T.}~\bibnamefont{Gu\'erin}, \bibfnamefont{T.}}
  \bibnamefont{and} \bibinfo{author}{\bibfnamefont{D.~S.} \bibnamefont{Dean}},
  \bibinfo{journal}{Phys. Rev. Lett.} \textbf{\bibinfo{volume}{115}},
  \bibinfo{pages}{020601} (\bibinfo{year}{2015}).

\bibitem[{\citenamefont{Barton}(1989)}]{barton1989elements}
\bibinfo{author}{\bibfnamefont{G.}~\bibnamefont{Barton}},
  \emph{\bibinfo{title}{Elements of Green's functions and propagation}}
  (\bibinfo{publisher}{Clarendon Press, Oxford}, \bibinfo{year}{1989}).

\bibitem[{\citenamefont{Carbonell and
  Whitaker}(1983)}]{carbonell1983dispersion}
\bibinfo{author}{\bibfnamefont{R.}~\bibnamefont{Carbonell}} \bibnamefont{and}
  \bibinfo{author}{\bibfnamefont{S.}~\bibnamefont{Whitaker}},
  \bibinfo{journal}{Chem. Eng. Sci.} \textbf{\bibinfo{volume}{38}},
  \bibinfo{pages}{1795} (\bibinfo{year}{1983}).

\bibitem[{\citenamefont{Drummond and Horgan}(1987)}]{drummond1987effective}
\bibinfo{author}{\bibfnamefont{I.}~\bibnamefont{Drummond}} \bibnamefont{and}
  \bibinfo{author}{\bibfnamefont{R.}~\bibnamefont{Horgan}},
  \bibinfo{journal}{J. Phys. A- Math. Gen.} \textbf{\bibinfo{volume}{20}},
  \bibinfo{pages}{4661} (\bibinfo{year}{1987}).

\bibitem[{\citenamefont{Dean and Oshanin}(2014)}]{dean2014approach}
\bibinfo{author}{\bibfnamefont{D.~S.} \bibnamefont{Dean}} \bibnamefont{and}
  \bibinfo{author}{\bibfnamefont{G.}~\bibnamefont{Oshanin}},
  \bibinfo{journal}{Phys. Rev. E} \textbf{\bibinfo{volume}{90}},
  \bibinfo{pages}{022112} (\bibinfo{year}{2014}).

\bibitem[{\citenamefont{Dean and Gu\'erin}(2014)}]{Dean2014PRE}
\bibinfo{author}{\bibfnamefont{D.~S.} \bibnamefont{Dean}} \bibnamefont{and}
  \bibinfo{author}{\bibfnamefont{T.}~\bibnamefont{Gu\'erin}},
  \bibinfo{journal}{Phys. Rev. E} \textbf{\bibinfo{volume}{90}},
  \bibinfo{pages}{062114} (\bibinfo{year}{2014}).

\bibitem[{\citenamefont{Stratonovich}(1958)}]{stratonovich1958oscillator}
\bibinfo{author}{\bibfnamefont{R.}~\bibnamefont{Stratonovich}},
  \bibinfo{journal}{Radiotekh Elektron. (Moscow)} \textbf{\bibinfo{volume}{3}},
  \bibinfo{pages}{497} (\bibinfo{year}{1958}).

\bibitem[{\citenamefont{Peters and Barenbrug}(2002)}]{Peters2002}
\bibinfo{author}{\bibfnamefont{E.}~\bibnamefont{Peters}} \bibnamefont{and}
  \bibinfo{author}{\bibfnamefont{T.}~\bibnamefont{Barenbrug}},
  \bibinfo{journal}{Phys. Rev. E} \textbf{\bibinfo{volume}{66}},
  \bibinfo{pages}{056701} (\bibinfo{year}{2002}).

\bibitem[{\citenamefont{Barenbrug et~al.}(2002)\citenamefont{Barenbrug, Peters,
  and Schieber}}]{Barenbrug2002}
\bibinfo{author}{\bibfnamefont{T.}~\bibnamefont{Barenbrug}},
  \bibinfo{author}{\bibfnamefont{E.}~\bibnamefont{Peters}}, \bibnamefont{and}
  \bibinfo{author}{\bibfnamefont{J.}~\bibnamefont{Schieber}},
  \bibinfo{journal}{J. Chem. Phys.} \textbf{\bibinfo{volume}{117}},
  \bibinfo{pages}{9202} (\bibinfo{year}{2002}).

\bibitem[{\citenamefont{Lamm and Schulten}(1983)}]{lamm1983extended}
\bibinfo{author}{\bibfnamefont{G.}~\bibnamefont{Lamm}} \bibnamefont{and}
  \bibinfo{author}{\bibfnamefont{K.}~\bibnamefont{Schulten}},
  \bibinfo{journal}{J. Chem. Phys.} \textbf{\bibinfo{volume}{78}},
  \bibinfo{pages}{2713} (\bibinfo{year}{1983}).

\bibitem[{\citenamefont{Dalle-Ferrier et~al.}(2011)\citenamefont{Dalle-Ferrier,
  Kr{\"u}ger, Hanes, Walta, Jenkins, and Egelhaaf}}]{DalleFerrier2011dynamics}
\bibinfo{author}{\bibfnamefont{C.}~\bibnamefont{Dalle-Ferrier}},
  \bibinfo{author}{\bibfnamefont{M.}~\bibnamefont{Kr{\"u}ger}},
  \bibinfo{author}{\bibfnamefont{R.~D.} \bibnamefont{Hanes}},
  \bibinfo{author}{\bibfnamefont{S.}~\bibnamefont{Walta}},
  \bibinfo{author}{\bibfnamefont{M.~C.} \bibnamefont{Jenkins}},
  \bibnamefont{and} \bibinfo{author}{\bibfnamefont{S.~U.}
  \bibnamefont{Egelhaaf}}, \bibinfo{journal}{Soft Matt.}
  \textbf{\bibinfo{volume}{7}}, \bibinfo{pages}{2064} (\bibinfo{year}{2011}).

\bibitem[{\citenamefont{Evstigneev et~al.}(2008)\citenamefont{Evstigneev,
  Zvyagolskaya, Bleil, Eichhorn, Bechinger, and Reimann}}]{Evstigneev2008}
\bibinfo{author}{\bibfnamefont{M.}~\bibnamefont{Evstigneev}},
  \bibinfo{author}{\bibfnamefont{O.}~\bibnamefont{Zvyagolskaya}},
  \bibinfo{author}{\bibfnamefont{S.}~\bibnamefont{Bleil}},
  \bibinfo{author}{\bibfnamefont{R.}~\bibnamefont{Eichhorn}},
  \bibinfo{author}{\bibfnamefont{C.}~\bibnamefont{Bechinger}},
  \bibnamefont{and} \bibinfo{author}{\bibfnamefont{P.}~\bibnamefont{Reimann}},
  \bibinfo{journal}{Phys. Rev. E} \textbf{\bibinfo{volume}{77}},
  \bibinfo{pages}{041107} (\bibinfo{year}{2008}).

\bibitem[{\citenamefont{Evers et~al.}(2013)\citenamefont{Evers, Hanes, Zunke,
  Capellmann, Bewerunge, Dalle-Ferrier, Jenkins, Ladadwa, Heuer,
  Casta{\~n}eda-Priego et~al.}}]{evers2013colloids}
\bibinfo{author}{\bibfnamefont{F.}~\bibnamefont{Evers}},
  \bibinfo{author}{\bibfnamefont{R.}~\bibnamefont{Hanes}},
  \bibinfo{author}{\bibfnamefont{C.}~\bibnamefont{Zunke}},
  \bibinfo{author}{\bibfnamefont{R.}~\bibnamefont{Capellmann}},
  \bibinfo{author}{\bibfnamefont{J.}~\bibnamefont{Bewerunge}},
  \bibinfo{author}{\bibfnamefont{C.}~\bibnamefont{Dalle-Ferrier}},
  \bibinfo{author}{\bibfnamefont{M.}~\bibnamefont{Jenkins}},
  \bibinfo{author}{\bibfnamefont{I.}~\bibnamefont{Ladadwa}},
  \bibinfo{author}{\bibfnamefont{A.}~\bibnamefont{Heuer}},
  \bibinfo{author}{\bibfnamefont{R.}~\bibnamefont{Casta{\~n}eda-Priego}},
  \bibnamefont{et~al.}, \bibinfo{journal}{Eur. Phys. J. Special Topics}
  \textbf{\bibinfo{volume}{222}}, \bibinfo{pages}{2995} (\bibinfo{year}{2013}).

\bibitem[{\citenamefont{Hanes et~al.}(2012)\citenamefont{Hanes, Dalle-Ferrier,
  Schmiedeberg, Jenkins, and Egelhaaf}}]{hanes2012colloids}
\bibinfo{author}{\bibfnamefont{R.~D.} \bibnamefont{Hanes}},
  \bibinfo{author}{\bibfnamefont{C.}~\bibnamefont{Dalle-Ferrier}},
  \bibinfo{author}{\bibfnamefont{M.}~\bibnamefont{Schmiedeberg}},
  \bibinfo{author}{\bibfnamefont{M.~C.} \bibnamefont{Jenkins}},
  \bibnamefont{and} \bibinfo{author}{\bibfnamefont{S.~U.}
  \bibnamefont{Egelhaaf}}, \bibinfo{journal}{Soft Matter}
  \textbf{\bibinfo{volume}{8}}, \bibinfo{pages}{2714} (\bibinfo{year}{2012}).

\bibitem[{\citenamefont{Speck and Seifert}(2006)}]{speck2006restoring}
\bibinfo{author}{\bibfnamefont{T.}~\bibnamefont{Speck}} \bibnamefont{and}
  \bibinfo{author}{\bibfnamefont{U.}~\bibnamefont{Seifert}},
  \bibinfo{journal}{Europhys. Lett.} \textbf{\bibinfo{volume}{74}},
  \bibinfo{pages}{391} (\bibinfo{year}{2006}).

\bibitem[{\citenamefont{Blickle et~al.}(2007)\citenamefont{Blickle, Speck,
  Lutz, Seifert, and Bechinger}}]{blickle2007einstein}
\bibinfo{author}{\bibfnamefont{V.}~\bibnamefont{Blickle}},
  \bibinfo{author}{\bibfnamefont{T.}~\bibnamefont{Speck}},
  \bibinfo{author}{\bibfnamefont{C.}~\bibnamefont{Lutz}},
  \bibinfo{author}{\bibfnamefont{U.}~\bibnamefont{Seifert}}, \bibnamefont{and}
  \bibinfo{author}{\bibfnamefont{C.}~\bibnamefont{Bechinger}},
  \bibinfo{journal}{Phys. Rev. Lett.} \textbf{\bibinfo{volume}{98}},
  \bibinfo{pages}{210601} (\bibinfo{year}{2007}).

\bibitem[{\citenamefont{Baiesi et~al.}(2011)\citenamefont{Baiesi, Maes, and
  Wynants}}]{baiesi2011modified}
\bibinfo{author}{\bibfnamefont{M.}~\bibnamefont{Baiesi}},
  \bibinfo{author}{\bibfnamefont{C.}~\bibnamefont{Maes}}, \bibnamefont{and}
  \bibinfo{author}{\bibfnamefont{B.}~\bibnamefont{Wynants}},
  \bibinfo{journal}{Proc. Roy. Soc. London A: Math. Phys.}
  \textbf{\bibinfo{volume}{467}}, \bibinfo{pages}{2792} (\bibinfo{year}{2011}).

\bibitem[{\citenamefont{Maes et~al.}(2013)\citenamefont{Maes, Safaverdi, Visco,
  and Van~Wijland}}]{maes2013fluctuation}
\bibinfo{author}{\bibfnamefont{C.}~\bibnamefont{Maes}},
  \bibinfo{author}{\bibfnamefont{S.}~\bibnamefont{Safaverdi}},
  \bibinfo{author}{\bibfnamefont{P.}~\bibnamefont{Visco}}, \bibnamefont{and}
  \bibinfo{author}{\bibfnamefont{F.}~\bibnamefont{Van~Wijland}},
  \bibinfo{journal}{Phys. Rev. E} \textbf{\bibinfo{volume}{87}},
  \bibinfo{pages}{022125} (\bibinfo{year}{2013}).

\end{thebibliography}

\end{document}